\documentclass[english]{article}
\usepackage[T1]{fontenc}
\usepackage[applemac]{inputenc}
\usepackage{geometry}
\usepackage{color}
\usepackage{babel}
\usepackage{mathtools}
\usepackage{amsmath}
\usepackage{amssymb}
\usepackage{esint}
\usepackage[unicode=true,pdfusetitle,
 bookmarks=true,bookmarksnumbered=false,bookmarksopen=false,
 breaklinks=false,pdfborder={0 0 1},backref=false,colorlinks=false]
 {hyperref}

\makeatletter
\usepackage{babel}

\makeatother

\begin{document}
\title{\textcolor{blue}{\huge{}A Suggested Answer To Wallstrom's Criticism:
Zitterbewegung Stochastic Mechanics I}}
\author{Maaneli Derakhshani\thanks{Email: m.derakhshani@uu.nl and maanelid@yahoo.com}\medskip{}
\medskip{}
\\
\emph{Department of Mathematics }\\
\emph{Utrecht University, Utrecht, The Netherlands}\thanks{Address when this work was initiated: Department of Physics \& Astronomy,
University of Nebraska-Lincoln, Lincoln, NE 68588, USA.} }
\maketitle
\begin{abstract}
Wallstrom's criticism of existing formulations of stochastic mechanics is that they fail to derive the empirical predictions of orthodox quantum mechanics because they require an ad hoc quantization condition on the postulated velocity potential, \emph{S}, in order to derive Schrödinger wave functions. We propose an answer to this criticism by modifying the Nelson-Yasue formulation of non-relativistic stochastic mechanics for a spinless particle with the following hypothesis: a spinless Nelson-Yasue particle of rest mass $m$ continuously undergoes a driven steady-state oscillation of `zitterbewegung' (\emph{zbw}) frequency, $\omega_{c}=\left(1/\hbar\right)mc^{2}$, in its instantaneous mean forward (and backward) translational rest frame. With this hypothesis we show that, in the lab frame, \emph{S} arises from imposing the constraint of conservative diffusions on the time-symmetrized steady-state phase of the \emph{zbw} particle, satisfies the required quantization condition, and evolves in time by the Hamilton-Jacobi-Madelung equations (when generalized to describe a statistical ensemble of \emph{zbw} particles). From the mathematical equivalence of Schrödinger's equation with the Hamilton-Jacobi-Madelung equations plus the quantization condition, Schrödinger wave functions for a spinless particle (in and excluding external fields) are thereby recovered. We also apply this `zitterbewegung stochastic mechanics' (ZSM) to the case of a central potential and show that it predicts angular momentum quantization. This paper sets the foundation for Part II, which will (primarily) work out the many-particle version of ZSM.\pagebreak{}

\tableofcontents{}
\end{abstract}

\section{Introduction}

Since its introduction by Fényes in 1952 \cite{Fenyes1952}, the goal
of the stochastic mechanics research program has been to derive quantum
theory from a classical-like statistical mechanics of particles undergoing
Brownian motion. Towards this end, non-relativistic and relativistic
models of stochastic mechanics have been constructed for both spin-0
particles \cite{Fenyes1952,Nelson1966,Nelson1967,Nelson1985,Yasue1978a,Yasue1979,Yasue1981,Yasue1981a,Yasue1977,Guerra1981,Guerra1983,Davidson1979,Davidson(2006),Nagasawa1996,Cufaro-Petroni1995,Cufaro-Petroni1997,Bacciagaluppi2003,Bacciagaluppi2012,Kobayashi2011,Zastawniak1990,Dohrn1978,Dohrn1979,Dohrn1985,Smolin1986,Smolin(2002),Markopoulou2004,Serva1988,Marra1990,Aldrovandi1990,Aldrovandi1992,Garbaczewski1992,Morato1995}
and spin-$1/2$ particles \cite{Dankel1970,Dohrn1979,Faris1982,Angelis1986,Garbaczewski1992}.
A non-relativistic theory of single-time and multi-time measurements
has also been developed \cite{Blanchard1986,Goldstein1987,Jibu1990,Blanchard1992,Peruzzi1996},
as have extensions of non-relativistic stochastic mechanics to finite
temperature and non-equilibrium open systems \cite{Yasue1978a,Roy2009,S.Bhattacharya2011,Kobayashi2011,Koide2013}.
Field theoretic generalizations also exist, for the cases of scalar
fields \cite{Guerra1973,Yasue1978,Davidson1980,Guerra1981,Koide2015a},
Maxwell fields \cite{Guerra1979,Koide2014}, vector-meson fields \cite{Siena1983},
the linearized gravitational field \cite{Davidson1982}, coupling
to dissipative environments \cite{Yasue1978,Lim1987}, non-Abelian
gauge theory \cite{Yasue1979}, bosonic string theory \cite{Santos99},
M-theory \cite{Smolin(2002)}, and background-independent quantum
gravity \cite{Markopoulou2004}. However, Wallstrom \cite{Wallstrom1989,Wallstrom1994}
pointed out that extant formulations of stochastic mechanics ultimately
fail to derive quantum mechanics because they require an ``ad hoc''
quantization condition on the postulated velocity potential, \textit{S},
in order to recover single-valued Schrödinger wave functions. Moreover,
this criticism appears to generalize to the field-theoretic and quantum
gravitational versions of stochastic mechanics developed before, during,
and after Wallstrom's publications, insofar as they require analogous
quantization conditions and don't seem to give non-circular justifications
for them.

Since Wallstrom, sporadic attempts have been made to answer his criticism
\cite{Carlen1989,Wallstrom1994,Bacciagaluppi2005,Smolin06,Fritsche2009,Catich2011,Schmelzer(2011),Groessing2011}.
However, in our view, all these attempts are either problematic or
limited in their applicability to stochastic mechanics (the follow
up paper, Part II, will give a discussion). Nevertheless, if a convincing
answer can be found, stochastic mechanics may once again be viewed
as a viable research program, and one that (in our view) offers elegant
solutions to many of the foundational problems with quantum mechanics.
As examples, stochastic mechanics would provide: (1) an unambiguous
solution to the quantum measurement problem (the local beables of
the theory on which measurement outcomes depend are point masses with
definite trajectories at all times) \cite{Goldstein1987,Jibu1990,Blanchard1992,Peruzzi1996};
(2) a novel and unambiguous physical interpretation of the wave function
(it is epistemic in the sense of being defined from field variables
describing a fictitious ensemble of point masses undergoing conservative
diffusions; and it has ontic properties in the specific sense that
the evolutions of said variables are constrained by beables over and
above the point masses) \cite{Nelson1985,Wallstrom1994,Derakhshani2016b};
(3) an explanation for why the position basis is preferred in decoherence
theory (the form of the Schrödinger Hamiltonian is a consequence of
the particle diffusion process happening in position space) \cite{Bacciagaluppi2005,Kobayashi2011};
and (4) a justification for the symmetry postulates for wave functions
of identical particles (they arise from natural symmetry conditions
on the particle trajectories, with the possibility of parastatistics
being excluded) \cite{Nelson1985,Goldstein1987,Bacciagaluppi2003}.

In this connection, it is worth mentioning that some of the aforementioned
virtues of stochastic mechanics, such as (1) and (4), are shared by
de Broglie-Bohm theories \cite{Holland1993,Bohm1995,Duerr2009,OriolsMompart2012,Goldstein2013,Oriols2016,Derakhshani2017b};
conversely, virtually all of the technical results obtained from de
Broglie-Bohm theories can be directly imported into stochastic mechanics
(basically because stochastic mechanics contains the dynamical equations
of de Broglie-Bohm theories as a subset).

This being said, stochastic mechanics (if viable) has a notably significant
difference from the `standard' approaches to interpreting or reformulating
or replacing the quantum formalism in a realist way that solves the
measurement problem, those being many-worlds theories \cite{Allori2009,Wallace2012,Vaidm2014},
de Broglie-Bohm theories \cite{Holland1993,Bohm1995,Duerr2009,OriolsMompart2012,Goldstein2013,Oriols2016},
and dynamical collapse theories \cite{Ghirardi2011,Adler2012,Bassi2013}.
In all these approaches, the wave function is interpreted as fundamental
and ontic (or as some kind of physical law \cite{GoldsteinZanghi2011,Goldstein2013,VassalloDeckertEsfeld2016}),
and the Schrödinger equation (or some nonlinear modification of it)
is taken as a dynamical law. So if stochastic mechanics succeeds in
deriving the Schrödinger equation and wave function, it constitutes
(arguably) the first example of a measurement-problem-free ontological
reconstruction of quantum mechanics in which the wave function could
be considered (in a well-defined sense) as genuinely derived and epistemic,
and the Schrödinger evolution as phenomenological rather than law-like\footnote{The recent ``Many-Interacting-Worlds'' (MIW) theory of Hall, Deckert,
and Wiseman \cite{Hall2014}, shares some of these features in that
it recovers the Schrödinger wave function as an effective, mean-field
description of a large number of real classical worlds interacting
through a non-classical (quantum) force. On the other hand, it seems
that their approach is also subject to Wallstrom's criticism in that
they also have to assume the quantization condition (or something
like it) on the dynamics of their classical worlds. Similar comments
apply to the ``Prodigal QM'' theory of Sebens \cite{Sebens2015}.

Similarly, the ``Trace Dynamics'' theory of Steven Adler \cite{Adler2002,Adler2012,Adler2013,Bassi2013}
aims to derive the quantum formalism as an approximation to the thermodynamic
limit of a statistical mechanical description of Grassmannian matrices
living on space-time. However, Trace Dynamics requires certain ad
hoc assumptions, namely that the state-vector in the thermodynamic
description has a norm-preserving nonlinear stochastic evolution.
Such an assumption is ad hoc because it seems to have no justification
from within the assumptions of Trace Dynamics, whereas it presumably
should have such a justification in order to sustain the claim that
Trace Dynamics derives the quantum formalism in a certain approximation.
(This view is also espoused by Bassi et al. in \cite{Bassi2013}.)
In this sense, it seems fair to say that the norm-perserving assumption
is to Trace Dynamics what the quantization condition is to (extant
formulations of) stochastic mechanics. }. Thus stochastic mechanics would (if viable) constitute a couterexample
to an implicit assumption that motivates the aforementioned standard
approaches - that the wave function and Schrödinger equation must
be part of the fundamental ontology (or laws) and dynamical laws,
respectively, in order to have a realist alternative to standard quantum
theory that solves the measurement problem, is empirically adequate,
and has a coherent physical/ontological interpretation.

It is also noteworthy that, as a dynamical theory of particle motion
in which probabilities play no fundamental role, stochastic mechanics
shares with de Broglie-Bohm theories the ability to justify the ``quantum
equilibrium'' density $|\psi|^{2}$ from typicality arguments \cite{Duerr1992}
and from dynamical relaxation of non-equilibrium densities to future
equilibrium \cite{Goldstein1987,Cufaro-Petroni1995,Cufaro-Petroni1997,Bacciagaluppi2012}.
As a result, stochastic mechanics can, on its own terms, be regarded
as a more general physical theory that contains quantum mechanics
as a fixed point - and outside this fixed point, it admits the possibility
of non-equilibrium physics, e.g., measurements more precise than the
uncertainty principle allows and superluminal signaling \cite{Nelson1985,Bohm1952I,Bohm1952II,Pearle2006}.
We will also argue in Part II \cite{Derakhshani2016b} that quantum
non-equilibrium states are more plausibly motivated in stochastic
mechanics than in deterministic de Broglie-Bohm theories.

For all these reasons and more, it seems worthwhile to consider whether
the central obstacle for the stochastic mechanics research program
- Wallstrom's criticism - can be surmounted. The objective of this
series of papers is to suggest how non-relativistic stochastic mechanics
for spinless particles can be modified to provide a non-ad-hoc physical
justification for the required quantization condition on $S$, and
thereby recover all and only the single-valued wave functions of non-relativistic
quantum mechanics. In this paper, we propose to modify the Nelson-Yasue
formulation \cite{Nelson1985,Yasue1981} of non-relativistic stochastic
mechanics for a spinless particle with the following hypothesis: a
spinless particle of rest mass, $m$, bounded to a harmonic potential
of natural frequency, $\omega_{c}=\left(1/\hbar\right)mc^{2}$, and
immersed in Nelson's hypothetical ether medium (appropriately modified
in its properties), undergoes a driven steady-state oscillation of
`zitterbewegung' (\emph{zbw}) frequency, $\omega_{c}$, in its instantaneous
mean forward (and backward) translational rest frame. With this hypothesis
we show that, in the lab frame, the stochastic mechanical velocity
potential, \emph{S}, arises from imposing the constraint of conservative
diffusions on the time-symmetrized steady-state phase of the \emph{zbw}
particle, implies the needed quantization condition, and evolves by
the stochastically derived Hamilton-Jacobi-Madelung equations (when
generalized to describe a statistical ensemble of \emph{zbw} particles).
This modification of Nelson-Yasue stochastic mechanics (NYSM), which
we term `zitterbewegung stochastic mechanics' (ZSM), then allows us
to derive the single-valued wave functions of non-relativistic quantum
mechanics for a spinless particle. The problem of justifying the quantization
condition is thereby reduced to justifying the zitterbewegung hypothesis.
Accordingly, it is among the tasks of Part II to argue that the hypothesis
can be justified in terms of physical/dynamical models and can be
plausibly generalized to particles with spin as well as relativistic
particles and fields.

The outline of this paper is as follows. In section 2, we give a concise
review of the formal derivation of the Schrödinger equation from NYSM
for a single, spinless particle in an external scalar potential. (Such
a review will be useful for the reader who is unfamiliar with NYSM,
and essential for following the logic and presentation of the arguments
later in the paper.) In section 3, we review the Wallstrom criticism.
In section 4, we introduce a classical model of a spinless zitterbewegung
particle which implies the quantization condition for the phase of
its oscillation, excluding and including interactions with external
fields. In each case, we extend the model to a classical Hamilton-Jacobi
statistical mechanics involving a Gibbsian ensemble of such particles,
with the purpose of making as clear as possible the physical assumptions
of the model in a well-established classical physics framework that
has conceptual and mathematical similarities to stochastic mechanics.
In section 5, we construct a Nelson-Yasue stochastic mechanics for
the zitterbewegung particle (ZSM), excluding and including field interactions.
In this way we derive one-particle Schrödinger equations with single-valued
wave functions that have (generally) multi-valued phases, and use
the hydrogen-like atom as a worked example.

This paper lays the foundation for Part II, where we will: (1) develop
the (non-trivial) many-particle cases of ZSM, (2) explicate the beables
of ZSM, (3) assess the plausibility and generalizability of the zitterbewegung
hypothesis, and (4) compare ZSM to other proposed answers to Wallstrom's
criticism.

\section{Nelson-Yasue Stochastic Mechanics}

In Edward Nelson's non-relativistic stochastic mechanics \cite{Nelson1966,Nelson1967,Nelson1985},
it is first hypothesized that the vacuum is pervaded by a homogeneous
and isotropic ``ether'' fluid with classical stochastic fluctuations
of uniform character. \footnote{The microscopic constituents of this ether are left unspecified by
Nelson; however, he suggests by tentative dimensional arguments relating
to the choice of diffusion constant in Eq. (3) (namely, that we can
write $\hbar=e^{2}/\alpha c$, where $\alpha$ is the fine-structure
constant and $e$ the elementary charge) that it may have an electromagnetic
origin \cite{Nelson1985}. } To ensure that observers in the ether can't distinguish absolute
rest from uniform motion, it is further hypothesized that the interaction
of a point mass with the ether is a frictionless diffusion process.
\footnote{Nelson points out \cite{Nelson1985} that this frictionless diffusion
process is an example of ``conservative diffusions'', or diffusions
in which the ensemble-averaged energy of the particle is conserved
in time (for a time-independent external potential). In other words,
on the (ensemble) average, there is no net transfer of energy between
the particle and the fluctuating ether, in contrast to classical Brownian
diffusions which are fundamentally dissipative in character.} Accordingly, a point particle of mass \emph{m} within this frictionless
ether will in general have its position 3-vector $\mathbf{q}(t)$
constantly undergoing diffusive motion with drift, as modeled by the
first-order stochastic differential equation, 
\begin{equation}
d\mathbf{q}(t)=\mathbf{b}(\mathbf{q}(t),t)dt+d\mathbf{W}(t).
\end{equation}
The vector $\mathbf{b}(\mathbf{q}(t),t)$ is the deterministic ``mean
forward'' drift velocity of the particle, and $\mathbf{W}(t)$ is
the Wiener process modeling the effect of the particle's interaction
with the fluctuating ether.

The Wiener increment, $d\mathbf{W}(t)$, is assumed to be Gaussian
with zero mean, independent of $d\mathbf{q}(s)$ for $s\leq t$, and
with covariance, 
\begin{equation}
\mathrm{E}_{t}\left[d\mathbf{W}_{i}(t)d\mathbf{W}_{j}(t)\right]=2\nu\delta_{ij}dt,
\end{equation}
where $\mathrm{E}_{t}$ denotes the conditional expectation at time
\emph{t}.

Note that although Equations (1-2) are formally the same as those
used for the kinematical description of classical Brownian motion
in the Einstein-Smoluchowski (ES) theory, the physical context is
different; the ES theory uses (1-2) to model the Brownian motion of
macroscopic particles in a classical fluid in the large friction limit
\cite{Nelson1967}, whereas Nelson uses (1-2) to model frictionless
stochastic motion (i.e., ``conservative diffusions'' \cite{Nelson1985})
for elementary particles interacting with a fluctuating ether fluid
that permeates the vacuum.

In this connection, it is further hypothesized that the magnitude
of the diffusion coefficient $\nu$ is proportional to the reduced
Planck's constant, and inversely proportional to the particle mass
\emph{m} so that 
\begin{equation}
\nu=\frac{\hbar}{2m}.
\end{equation}

In addition to (1), the particle's trajectory $\mathbf{q}(t)$ can
also satisfy the time-reversed equation 
\begin{equation}
d\mathbf{q}(t)=\mathbf{b}_{*}(\mathbf{q}(t),t)dt+d\mathbf{W}_{*}(t),
\end{equation}
where $\mathbf{b}_{*}(\mathbf{q}(t),t)$ is the mean backward drift
velocity, and $d\mathbf{W}_{*}(t)=d\mathbf{W}(-t)$ is the backward
Wiener process. The $d\mathbf{W}_{*}(t)$ has all the properties of
$d\mathbf{W}(t)$, except that it is independent of $d\mathbf{q}(s)$
for $s\geq t$. With these conditions on $d\mathbf{W}(t)$ and $d\mathbf{W}_{*}(t)$,
(1) and (4) respectively define forward and backward Markov processes
on $\mathbb{R}^{3}$.

The forwards and backwards transition probabilities defined by (1)
and (4), respectively, should be understood, in some sense, as ontic
probabilities \cite{Uffink2006,BacciagaluppiProbab2011}. (Generally
speaking, `ontic probabilities' can be understood as probabilities
about objective physical properties of the \emph{N}-particle system,
as opposed to `epistemic probabilities' \cite{Arntzenius1995} which
are about our ignorance of objective physical properties of the \emph{N}-particle
system.) Just how `ontic' these transition probabilities should be
is an open question. One possibility is that these transition probabilities
should be viewed as phenomenologically modeling complicated deterministic
interactions of a massive particle (or particles) with the fluctuating
ether, in analogy with how equations such as (1) and (4) are used
in the ES to phenomenologically model the complicated deterministic
interactions of a macroscopic particle immersed in a fluctuating classical
fluid of finite temperature \cite{Nelson1967}. Another possibility
is that the fluctuations of the ether are irreducibly stochastic,
and this irreducible stochasticity is 'transferred' to a particle
immersed in and interacting with the ether. We prefer the former possibility,
but acknowledge that the latter possibility is also viable. \footnote{Concerning whether or not the forward and backwards transition probabilities
should be understood as `objective' (i.e., as chances governed by
natural law) versus `subjective' (i.e., encoding our expectations
or degrees of belief) \cite{FriggHoefer2010,Glynn2010,Emery2015},
this seems to depend on whether the transition probabilities are merely
phenomenological (in which case they would seem to be subjective)
or reflect irreducible stochasticity in the ether (in which case they
would seem to be objective). Our preference for viewing the transition
probabilities as phenomenological seems to commit us to the subjective
view, but the objective view also seems viable (the objective view
is taken by Bacciagaluppi in \cite{Bacciagaluppi2005,Bacciagaluppi2012}).
It is worth noting that, under the objective view, the backwards transition
probabilities can be regarded as being just as objective/law-like
as the forwards transition probabilities (but see \cite{Arntzenius1995}
for a different view).}

Associated to the trajectory $\mathbf{q}(t)$ is the probability density
$\rho(\mathbf{q},t)=n(\mathbf{q},t)/N$, where $n(\mathbf{q},t)$
is the number of particles per unit volume and $N$ is the total number
of particles in a definite region of space. Corresponding to (1) and
(4), then, are the forward and backward Fokker-Planck equations, 
\begin{equation}
\frac{\partial\rho(\mathbf{q},t)}{\partial t}=-\nabla\cdot\left[\mathbf{b(}\mathbf{q},t)\rho(\mathbf{q},t)\right]+\frac{\hbar}{2m}\nabla^{2}\rho(\mathbf{q},t),
\end{equation}
and 
\begin{equation}
\frac{\partial\rho(\mathbf{q},t)}{\partial t}=-\nabla\cdot\left[\mathbf{b}_{*}(\mathbf{q},t)\rho(\mathbf{q},t)\right]-\frac{\hbar}{2m}\nabla^{2}\rho(\mathbf{q},t),
\end{equation}
where we require that $\rho(\mathbf{q},t)$ satisfies the normalization
condition, 
\begin{equation}
\int\rho_{0}(\mathbf{q})d^{3}q=1.
\end{equation}

We emphasize that, in contrast to the transition probabilities defined
by (1) and (4), the probability distributions satisfying (5) and (6)
are epistemic distributions in the sense that they are distributions
over a Gibbsian ensemble of identical systems (i.e., the distributions
reflect our ignorance of the actual positions of the particles). Nevertheless,
for an epistemic distribution satisfying (5) or (6) at time $t$,
its subsequent evolution will be determined by the ontic transition
probabilities so that the distribution at later times will partly
come to reflect ontic features of the \emph{N}-particle system, and
may asymptotically become independent of the initial distribution.
\footnote{I thank Guido Bacciagaluppi for emphasizing this point.}
Of course, the asymptotic distribution would still be epistemic in
the sense of encoding our ignorance of the actual particle positions,
even though it would be determined by the ontic features of the system.

A frictionless (hence energy-conserving or conservative) diffusion
process such as Nelson's should have a time-symmetric probability
density evolution. The Fokker-Planck equations (5-6), on the other
hand, describe time-asymmetric evolutions in opposite time directions.
The reason is that, given all possible solutions to (1), one can define
as many forward processes as there are possible initial distributions
satisfying (5); likewise, given all possible solutions to (4), one
can define as many backward processes as there are possible `initial'
distributions satisfying (6). Consequently, the forward and backward
processes are both underdetermined, and neither (1) nor (4) has a
well-defined time-reversal. We must therefore restrict the diffusion
process to simultaneous solutions of (5) and (6).

Note that the sum of (5) and (6) gives the continuity equation 
\begin{equation}
\frac{\partial\rho({\normalcolor \mathbf{q}},t)}{\partial t}=-\nabla\cdot\left[\mathbf{v}(\mathbf{q},t)\rho(\mathbf{q},t)\right],
\end{equation}
where 
\begin{equation}
\mathbf{v}(\mathbf{q},t)\coloneqq\frac{1}{2}\left[\mathbf{b}(\mathbf{q},t)+\mathbf{b}_{*}(\mathbf{q},t)\right]
\end{equation}
is called the ``current velocity'' field. As it stands, this current
velocity field could have vorticity. But if vorticity is allowed,
then the time-reversal operation on (5.8) will change the orientation
of the curl, thus distinguishing time directions \cite{PenaCetto1982,CaliariInversoMorato2004,Bacciagaluppi2012}.
So we impose 
\begin{equation}
\mathbf{v}(\mathbf{q}.t)=\frac{\nabla S(\mathbf{q},t)}{m},
\end{equation}
or that the current velocity field is irrotational. Accordingly, (8)
becomes 
\begin{equation}
\frac{\partial\rho({\normalcolor \mathbf{q}},t)}{\partial t}=-\nabla\cdot\left[\frac{\nabla S(\mathbf{q},t)}{m}\rho(\mathbf{q},t)\right],
\end{equation}
a time-reversal invariant evolution equation for the single-time density
$\rho({\normalcolor \mathbf{q}},t)$.

Physically speaking, the \textit{S} function in (10-11) has the interpretation
of a velocity potential connected with a Gibbsian ensemble of fictitious,
non-interacting, identical particles with density $\rho(\mathbf{q},t)$,
where each particle in the ensemble differs from the other in its
initial position (hence the dependence of $S$ on the generalized
coordinate $\mathbf{q}$) and initial irrotational velocity given
by (10). \footnote{Of course, one can still add to $\nabla S$ a solenoidal vector field
of any magnitude and, upon insertion into (8), recover the same continuity
equation \cite{Bacciagaluppi1999,HollandPhilipp2003}. But the assumption
of only irrotational flow velocity is the simplest one, and as we
already mentioned, it follows from the requirement of time symmetry
for the $\rho({\normalcolor \mathbf{q}},t)$ of the diffusion process. } It is thereby analogous to the \textit{S} function in the Hamilton-Jacobi
formulation of classical statistical mechanics for a single point
particle \cite{Schiller1962,Rosen1964,Holland1993,Ghose2002,Nikolic2006,Nikolic2007}.

Note also that subtracting (5) from (6) yields equality on the right
hand side of 
\begin{equation}
\mathbf{u}(\mathbf{q},t)\coloneqq\frac{1}{2}\left[\mathbf{b}(\mathbf{q},t)-\mathbf{b}_{*}(\mathbf{q},t)\right]=\frac{\hbar}{2m}\frac{\nabla\rho(\mathbf{q},t)}{\rho(\mathbf{q},t)},
\end{equation}
where $\mathbf{u}(\mathbf{q},t)$ is called the ``osmotic velocity''
field (because it has the same dependence on the density as the velocity
acquired by a classical Brownian particle in equilibrium with respect
to an external force, in the ES theory \cite{Nelson1966,Nelson1967,Nelson1985}).

As a consequence of (9), (10), and (12), we have that $\mathbf{b}=\mathbf{v}+\mathbf{u}$
and $\mathbf{b}_{*}=\mathbf{v}-\mathbf{u}$, which when inserted back
into (5) and (6), respectively, reduce both Fokker-Planck equations
to the time-reversal invariant continuity equation (11). So the combination
of (9), (10), and (12) fixes $\rho$ as the common, single-time, `equilibrium'
probability density (in analogy with a thermal equilibrium density)
for solutions of (1) and (4), even though it is a time-dependent density.

In our view, the physical meaning of (12) has been misconstrued by
some researchers \cite{Bohm1989,Kyprianidis1992,Smolin06,Smolin2012}
to imply that $\rho$ must be interpreted as the physical cause of
the osmotic velocity of Nelson's particle. We want to stress that
this is not the case, and that such an interpretation would be logically
and physically inconsistent with the definition of $\rho$ as a probability
density. Instead, Nelson physically motivates his osmotic velocity
by analogy with the osmotic velocity in the ES theory \cite{Nelson1966,Nelson1967}
- essentially, he postulates the presence of an external (i.e., not
sourced by the particle) potential, $U(\mathbf{q},t)$, which couples
to the particle via some coupling constant, $\mu$, such that $R(\mathbf{q}(t),t)=\mu U(\mathbf{q}(t),t)$
defines a `potential momentum' for the particle. \footnote{It should be emphasized that $U(\mathbf{q},t)$ is not defined over
an ensemble of systems, but is a real physical field on 3-space analogous
to the classical external potential, $V(\mathbf{q},t)$, that causes
the osmotic velocity of a Brownian particle in the E-S theory. Nelson
does not specify whether $U(\mathbf{q},t)$ is sourced by the ether
or is an independently existing field on space-time, nor does he specify
whether the coupling $\mu$ corresponds to any of the fundamental
force interactions of the Standard Model. These elements of his theory
are phenomenological hypotheses that presumably should be made more
precise in a `deeper' extension of stochastic mechanics. Nonetheless,
as we will see in Part II, the many-particle extension of stochastic
mechanics puts additional constraints on how the osmotic potential
should be understood.} (Hereafter we shall permit ourselves to refer to $U(\mathbf{q},t)$
and $R(\mathbf{q},t)$ interchangeably as the `osmotic potential'.)
When $U(\mathbf{q},t)$ is spatially varying, it imparts to the particle
a momentum, $\nabla R(\mathbf{q},t)|_{\mathbf{q}=\mathbf{q}(t)}$,
which is then counter-balanced by the ether fluid's osmotic impulse
pressure, $\left(\hbar/2m\right)\nabla\ln[n(\mathbf{q},t)]|_{\mathbf{q}=\mathbf{q}(t)}$.
This leads to the equilibrium condition $\nabla R/m=\left(\hbar/2m\right)\nabla\rho/\rho$
(using $\rho=n/N$), which implies that $\rho$ depends on $R$ as
$\rho=e^{2R/\hbar}$ for all times. Hence, the \textit{physical cause}
of $\mathbf{u}$ is $R$ (or technically \emph{U}), and (12) is just
a mathematically equivalent and convenient rewriting of this relation.

So far our discussion has been restricted to the first-order stochastic
differential equations for Nelson's particle, and the associated Fokker-Planck
evolutions. In order to discuss the second-order dynamics for Nelson's
particle, we must first motivate Nelson's analogues of the Ornstein-Uhlenbeck
mean derivatives. In the Itô calculus, the mean forward and backward
derivatives of a solution $\mathbf{q}(t)$ satisfying (1) and (4)
are respectively defined as 
\begin{equation}
D\mathbf{q}(t)=\underset{_{\Delta t\rightarrow0^{+}}}{lim}\mathrm{E_{t}}\left[\frac{q(t+\Delta t)-q(t)}{\Delta t}\right],
\end{equation}
and 
\begin{equation}
D_{*}\mathbf{q}(t)=\underset{_{\Delta t\rightarrow0^{+}}}{lim}\mathrm{E_{t}}\left[\frac{q(t)-q(t-\Delta t)}{\Delta t}\right].
\end{equation}
Because $d\mathbf{W}(t)$ and $d\mathbf{W}_{*}(t)$ are Gaussian with
zero mean, it follows that $D\mathbf{q}(t)=\mathbf{b}(\mathbf{q}(t),t)$
and $D_{*}\mathbf{q}(t)=\mathbf{b}_{*}(\mathbf{q}(t),t)$. To compute
the second mean derivative, $D\mathbf{b}(\mathbf{q}(t),t)$ (or $D_{*}\mathbf{b}(\mathbf{q}(t),t)$),
we must expand $\mathbf{b}$ in a Taylor series up to terms of order
two in $d\mathbf{q}(t)$: 
\begin{equation}
d\mathbf{b}(\mathbf{q}(t),t)=\frac{\partial\mathbf{b}(\mathbf{q}(t),t)}{\partial t}dt+d\mathbf{q}(t)\cdot\nabla\mathbf{b}(\mathbf{q}(t),t)+\frac{1}{2}\underset{i,j}{\sum}d\mathbf{\mathit{q}}_{i}(t)d\mathbf{\mathit{q}}_{j}(t)\frac{\partial^{2}\mathbf{b}(\mathbf{q}(t),t)}{\partial\mathbf{\mathit{q}}_{i}\partial\mathit{q}_{j}}+\ldots.
\end{equation}
From (1), we can replace $dx_{i}(t)$ by $dW_{i}(t)$ in the last
term, and when taking the conditional expectation in (13), we can
replace $d\mathbf{q}(t)\cdot\nabla\mathbf{b}(\mathbf{q}(t),t)$ by
$\mathbf{b}(\mathbf{q}(t),t)\cdot\nabla\mathbf{b}(\mathbf{q}(t),t)$
since $d\mathbf{W}(t)$ is independent of $\mathbf{q}(t)$ and has
mean 0. Using (2-3), we then obtain 
\begin{equation}
D\mathbf{b}(\mathbf{q}(t),t)=\left[\frac{\partial}{\partial t}+\mathbf{b}(\mathbf{q}(t),t)\cdot\nabla+\frac{\hbar}{2m}\nabla^{2}\right]\mathbf{b}(\mathbf{q}(t),t),
\end{equation}
and likewise 
\begin{equation}
D_{*}\mathbf{b}_{*}(\mathbf{q}(t),t)=\left[\frac{\partial}{\partial t}+\mathbf{b}_{*}(\mathbf{q}(t),t)\cdot\nabla-\frac{\hbar}{2m}\nabla^{2}\right]\mathbf{b}_{*}(\mathbf{q}(t),t).
\end{equation}
Using (16-17), along with Newton's 2nd law, Nelson wanted to construct
an expression for the `mean acceleration' of the particle consistent
with the principle of time-symmetry. He proposed 
\begin{equation}
m\mathbf{a}(\mathbf{q}(t),t)=\frac{m}{2}\left[D_{*}D+DD_{*}\right]\mathbf{q}(t)=-\nabla V(\mathbf{q},t)|_{\mathbf{q}=\mathbf{q}(t)}.
\end{equation}
Physically, this equation says that the mean acceleration Nelson's
particle feels in the presence of an external (conservative) force
is the equal-weighted average of its mean forward drift $\mathbf{b}$
transported backwards in time, with its mean backward drift $\mathbf{b}_{*}$
transported forwards in time. It is this peculiar mean dynamics that
preserves the time-symmetry of Nelson's diffusion process.

Of course, other time-symmetric mean accelerations are possible. For
example, $(1/2)[D^{2}+D_{*}^{2}]\mathbf{q}(t)$, or any weighted average
of this with (18). So it may be asked: what other physical principles
(if any) privilege Nelson's choice? As first shown by Yasue \cite{Yasue1981,Yasue1981a}
and later adopted by Nelson \cite{Nelson1985}, a physically well-motivated
stochastic variational principle can give (18). \footnote{Another widely used stochastic variational principle is the one due
to Guerra and Morato \cite{Guerra1983}. We don't use their approach
because it entails an $S$ function that's globally single-valued,
which excludes the possibility of systems with angular momentum \cite{Wallstrom1989}
and therefore will not be applicable to our proposed answer to Wallstrom's
criticism.} Consider the ensemble-averaged, time-symmetric mean action 
\begin{equation}
\begin{aligned}J & =\mathrm{E}\left[\int_{t_{i}}^{t_{f}}\left\{ \frac{1}{2}\left[\frac{1}{2}m\mathbf{b}(\mathbf{q}(t),t)^{2}+\frac{1}{2}m\mathbf{b}_{*}(\mathbf{q}(t),t)^{2}\right]-V(\mathbf{q}(t),t)\right\} dt\right]\\
 & =\mathrm{E}\left[\int_{t_{i}}^{t_{f}}\left\{ \frac{1}{2}m\mathbf{v}^{2}+\frac{1}{2}m\mathbf{u}^{2}-V\right\} dt\right].
\end{aligned}
\end{equation}
In other words, for a particle in a (possibly) time-dependent potential
$V$, undergoing the Markov processes given by (1) and (4) with the
restriction to simultaneous solutions of the Fokker-Planck equations
via (9), (10), and (12), a time-symmetric mean Lagrangian can be defined
by averaging together the mean Lagrangians associated with the forward
and backward processes. The \textcolor{black}{ensemble averaged action
obtained from} this time-symmetric mean Lagrangian then corresponds
to (19), where $\mathrm{E}\left[...\right]$ denotes the absolute
expectation. Upon imposing the conservative diffusion condition through
the variational principle, 
\begin{equation}
J=\mathrm{E}\left[\int_{t_{i}}^{t_{f}}\left\{ \frac{1}{2}m\mathbf{v}^{2}+\frac{1}{2}m\mathbf{u}^{2}-V\right\} dt\right]=extremal,
\end{equation}
a straightforward computation (see Appendix A) shows that this implies
(18) as the equation of motion. If, instead, we had allowed the mean
kinetic energy terms in (19) to not be positive-definite and used
the alternative time-symmetric mean kinetic energy, $(1/2)m\mathbf{b}\mathbf{b}_{*}=(1/2)m(\mathbf{v}^{2}-\mathbf{u}^{2})$,
then it can be shown \cite{Hasegaw1986,Kyprianidis1992,Davidson(2006)}
that imposing (20) would give the alternative time-symmetric mean
acceleration involving the derivatives $[D^{2}+D_{*}^{2}]$. \footnote{Additionally, Davidson \cite{Davidson(2006)} showed that by defining
a Lagrangian of the form $(1/2)m\left[(1/2)\left(\mathbf{b}^{2}+\mathbf{b}_{*}^{2}\right)-(\beta/8)(\mathbf{b}-\mathbf{b}_{*})^{2}\right]$,
where $\beta$ is a constant, the resulting equation of motion is
also equivalent to the usual Schrödinger equation, provided that the
diffusion coefficient $\nu=(1/\sqrt{1-\beta/2})\frac{\hbar}{2m}$.
We can see, however, that our criterion of restricting the kinetic
energy terms in the Lagrangian to only terms that are positive-definite,
excludes Davidson's choice of Lagrangian too.} So Nelson's mean acceleration choice is justified by the principle
of time-symmetry \textit{and} the natural physical requirement that
all the contributions to the mean kinetic energy of the Nelsonian
particle should be positive-definite.

By applying the mean derivatives in (18) to $\mathbf{q}(t)$, using
that $\mathbf{b}=\mathbf{v}+\mathbf{u}$ and $\mathbf{b}_{*}=\mathbf{v}-\mathbf{u}$,
and removing the dependence of the mean acceleration on the actual
particle trajectory $\mathbf{q}(t)$ so that $\mathbf{a}(\mathbf{q}(t),t)$
gets replaced by the mean acceleration field $\mathbf{a}(\mathbf{q},t)$,
a straightforward computation gives 
\begin{equation}
\begin{aligned}m\mathbf{a}(\mathbf{q},t) & =m\left[\frac{\partial\mathbf{v}(\mathbf{q},t)}{\partial t}+\mathbf{v}(\mathbf{q},t)\cdot\nabla\mathbf{v}(\mathbf{q},t)-\mathbf{u}(\mathbf{q},t)\cdot\nabla\mathbf{u}(\mathbf{q},t)-\frac{\hbar}{2m}\nabla^{2}\mathbf{u}(\mathbf{q},t)\right]\\
 & =\nabla\left[\frac{\partial S(\mathbf{q},t)}{\partial t}+\frac{\left(\nabla S(\mathbf{q},t)\right)^{2}}{2m}-\frac{\hbar^{2}}{2m}\frac{\nabla^{2}\sqrt{\rho(\mathbf{q},t)}}{\sqrt{\rho(\mathbf{q},t)}}\right]=-\nabla V(\mathbf{q},t).
\end{aligned}
\end{equation}
The mean acceleration field $\mathbf{a}(\mathbf{q},t)$ describes
the possible mean accelerations of the actual particle given all of
the possible spatial locations that the actual particle can occupy
at time \emph{t}. In other words, $\mathbf{a}(\mathbf{q},t)$ is the
mean acceleration field connected with the set of fictitious particles
forming the Gibbsian ensemble that reflects our ignorance of the actual
trajectory $\mathbf{q}(t)$ \cite{Holland1993}. Integrating both
sides of (21), and setting the arbitrary integration constants equal
to zero, we then obtain the Quantum Hamilton-Jacobi equation, 
\begin{equation}
-\frac{\partial S(\mathbf{q},t)}{\partial t}=\frac{\left(\nabla S(\mathbf{q},t)\right)^{2}}{2m}+V(\mathbf{q},t)-\frac{\hbar^{2}}{2m}\frac{\nabla^{2}\sqrt{\rho(\mathbf{q},t)}}{\sqrt{\rho(\mathbf{q},t)}},
\end{equation}
which describes the total energy field over the possible positions
of the actual\emph{ }point mass, and upon evaluation at $\mathbf{q}=\mathbf{q}(t),$
the total energy of the point mass along its actual trajectory.

Although the last term on the right hand side of (22) is often called
the ``quantum potential'', we note that it arises here from the
osmotic kinetic energy term in (19). So the quantum potential must
be physically understood in stochastic mechanics as a kinetic energy
field (which hereafter we prefer to call the `quantum kinetic' for
accuracy of meaning) arising from the osmotic velocity field.

The pair of nonlinear equations coupling the evolution of $\rho$
and $S$, as given by (11) and (22), are generally known as the Hamilton-Jacobi-Madelung
(HJM) equations, and can be formally identified with the imaginary
and real parts of the Schrödinger equation under polar decomposition
\cite{Takabayasi1952,Holland1993}. Therefore, (11) and (22) can be
formally rewritten as the Schrödinger equation, 
\begin{equation}
i\hbar\frac{\partial\psi(\mathbf{q},t)}{\partial t}=-\frac{\hbar^{2}}{2m}\nabla^{2}\psi(\mathbf{q},t)+V(\mathbf{q},t)\psi(\mathbf{q},t),
\end{equation}
where $\psi(\mathbf{q},t)=\sqrt{\rho(\mathbf{q},t)}e^{iS(\mathbf{q},t)/\hbar}$.
In contrast to other ontological formulations of quantum mechanics,
this wave function must be interpreted as an epistemic field in the
sense that it encodes information about the possible position and
momenta states that the actual particle can occupy at any instant,
since it is defined in terms of the ensemble variables $\rho$ and
$S$. \footnote{Though it may not be obvious here, this interpretation of the Nelson-Yasue
wave function is not undermined by the Pusey-Barrett-Rudolph theorem
\cite{Pusey2012}. Whereas this theorem assumes factorizability/separability
of the ``ontic state space'', the ontic osmotic potential, $U$,
which is encoded in the amplitude of the wave function via $R$ and
plays a role in the particle dynamics via (21), is in general not
separable when extended to the \emph{N}-particle case (as will be
shown in Part II \cite{Derakhshani2016b}). } Although the treatment here did not include coupling to electromagnetic
potentials, it is straightforward to do so \cite{Nelson1985} (see
also Appendix A ).

\section{Wallstrom's Criticism}

In the previous section, we referred to the correspondence between
the HJM equations and (23) as only formal because we had not considered
the boundary conditions that must be imposed on solutions of the Schrödinger
equation and the HJM equations, respectively, in order for mathematical
equivalence to be established. In standard quantum mechanics, it is
well-known that physical wave functions satisfying the Schrödinger
equation are required to be single-valued. For the HJM equations,
it was not specified in the Nelson-Yasue derivation whether \textit{S}
is assumed to be single-valued, arbitrarily multi-valued, or multi-valued
in accordance with a quantization condition. Wallstrom \cite{Wallstrom1989,Wallstrom1994}
showed that for all existing formulations of stochastic mechanics,
all these possible conditions on $S$ are problematic in one way or
another.

If $S$ is constrained to be single-valued, then stochastic mechanical
theories exclude single-valued Schrödinger wave functions with angular
momentum. This is so because single-valued wave functions with angular
momentum have phase factors of the form $\exp\left(i\mathrm{m}\varphi\right),$
where $\mathrm{m}$ is an integer and $\varphi$ is the azimuthal
angle, which implies that $S(\varphi)=\mathrm{m}\hbar\varphi$. By
contrast, if $S$ is assumed to be arbitrarily multi-valued, they
produce all the single-valued wave functions of the Schrödinger equation,
along with infinitely many multi-valued `wave functions', which smoothly
interpolate between the single-valued wave functions. This can be
seen by comparing solutions of the Schrödinger and HJM equations for
a two-dimensional central potential, $V(\mathbf{r})$ \cite{Wallstrom1989}.
The Schrödinger equation with $V(\mathbf{r})$ has single-valued wave
functions of the form $\psi_{\mathrm{m}}(\mathbf{r},\varphi)=R_{\mathrm{m}}(\mathrm{\mathbf{r}})exp(i\mathrm{m}\varphi),$
where $\psi_{\mathrm{m}}(\mathrm{\mathbf{r}},\varphi)=\psi_{\mathrm{m}}(\mathrm{\mathbf{r}},\varphi+2\pi n)$,
implying that $\mathrm{m}$ is an integer. For the HJM equations,
however, the solutions $\rho_{\mathrm{m}}=|R_{\mathrm{m}}(\mathrm{\mathbf{r}})|^{2}$
and $\boldsymbol{\mathbf{\mathrm{v}}}_{\mathrm{m}}=\left(\mathrm{m}\hbar/mr\right)\hat{\varphi}$
don't require $\mathrm{m}$ to be integral. To see this, consider
the effective central potential, $V_{a}(\boldsymbol{\mathrm{r}})=V(\boldsymbol{\mathrm{r}})+a/r^{2}$,
where $a$ is a positive real constant. For this potential, consider
the Schrödinger equation with stationary solution $\psi_{a}(\mathbf{r},\varphi)=R_{\mathrm{a}}(\mathrm{\mathbf{r}})exp(i\varphi)$,
where $\mathrm{m}=1$ and radial component corresponding to the ground
state solution of the radial equation. This wave function yields osmotic
and current velocities, $\mathbf{u}{}_{a}$ and $\mathrm{\mathbf{v}}{}_{a}$,
which satisfy (11) and (21) with the potential $V_{a}$: 
\begin{equation}
0=\frac{\partial\rho{}_{a}}{\partial t}=-\nabla\cdot\left(\mathbf{v}{}_{a}\rho{}_{a}\right),
\end{equation}
\begin{equation}
0=\frac{\partial\mathbf{v}{}_{a}}{\partial t}=-\nabla\left(V+\frac{a}{r^{2}}\right)-\mathbf{v}{}_{a}\cdot\nabla\mathbf{v}{}_{a}+\mathbf{u}{}_{a}\cdot\nabla\mathbf{u}{}_{a}+\frac{\hbar^{2}}{2m}\nabla^{2}\mathbf{u}{}_{a}.
\end{equation}
Using $\mathbf{v}{}_{a}=\left(\hbar/mr\right)\hat{\varphi}$ and $\mathbf{v}{}_{a}\cdot\nabla\mathbf{v}{}_{a}=\nabla\left[m\mathbf{v}{}_{a}^{2}/2\right]$,
we can then rewrite (25) as 
\begin{equation}
\begin{aligned}0 & =-\nabla V-\nabla\left(\frac{a}{r^{2}}+\frac{1}{2}m\mathbf{v}{}_{a}^{2}\right)+\mathbf{u}{}_{a}\cdot\nabla\mathbf{u}{}_{a}+\frac{\hbar^{2}}{2m}\nabla^{2}\mathbf{u}{}_{a}\\
 & =-\nabla V-\frac{m}{2}\nabla\left(\frac{2ma}{\hbar^{2}}+1\right)\mathbf{v}{}_{a}^{2}+\mathbf{u}{}_{a}\cdot\nabla\mathbf{u}{}_{a}+\frac{\hbar^{2}}{2m}\nabla^{2}\mathbf{u}{}_{a}.
\end{aligned}
\end{equation}
This gives us $\mathbf{v}'_{a}=\mathbf{v}{}_{a}\sqrt{2ma/\hbar^{2}+1}$
and $\mathbf{u}'_{a}=\mathbf{u}{}_{a}$. Note that since $a$ is a
constant that can take any positive real value, $\mathbf{v}'_{a}$
is not quantized, and yet it is a solution of the HJM equations. By
contrast, in the quantum mechanical version of this problem, we would
have $V_{a}(\boldsymbol{\mathrm{r}})=V(\boldsymbol{\mathrm{r}})+\mathrm{m}^{2}/2r^{2}$,
where $\mathrm{m}=\sqrt{2ma/\hbar^{2}+1}$ would be integral due to
the single-valuedness condition on $\psi_{\mathrm{m}}$. In other
words, the $\mathbf{v}{}_{a}$ and $\mathbf{u}{}_{a}$ in stochastic
mechanics only correspond to a single-valued wave function when $a$
is an integer, and this is true of all systems of two dimensions or
higher. Equivalently, we may say that the HJM equations predict a
continuum of energy and momentum states for the particle, which smoothly
interpolate between the quantized energy and momentum eigenvalues
predicted by the quantum mechanical case. \footnote{Before Wallstrom's critiques, it was pointed out by Albeverio and
Hoegh-Krohn \cite{Albeverio1974} as well as Goldstein \cite{Goldstein1987}
that, for the cases of stationary bound states with nodal surfaces
that separate the manifold of diffusion into disjoint components,
Nelson's equations (the HJM equations and his stochastic differential
equations) contain more solutions than Schrödinger 's equation. In
addition, Goldstein \cite{Goldstein1987} was the first to point out
that solutions exist to the HJM equations which don't correspond to
any single-valued solution of the Schrödinger equation, for the case
of a multiply-connected configuration space. Nevertheless, Wallstrom's
example of extraneous solutions is of a more general nature, as it
applies to a simply-connected space where the diffusion process is
not separated into disjoint components.}

The only condition on $S$ (and hence the current velocity $\mathbf{v}{}_{a}$)
that allows stochastic mechanics to recover all and only the single-valued
wave functions of the Schrödinger equation is the condition that the
change in $S$ around any closed loop $L$ in space (with time held
constant) is equal to an integer multiple of Planck's constant, \footnote{Wallstrom notes that Takabayasi \cite{Takabayasi1952} was first to
recognize the necessity of this quantization condition and suggests
{[}private communication{]} that priority of credit for this discovery
should go to him \cite{Wallstrom1994}. However, it seems that Takabayasi
only recognized this issue in the context of Bohm's 1952 hidden-variables
theory, even though Fényes proposed the first formulation of stochastic
mechanics that same year \cite{Fenyes1952}. Wallstrom appears to
have been the first in the literature to recognize and discuss the
full extent of this inequivalence in the context of stochastic mechanical
theories.} or 
\begin{equation}
\oint_{L}dS=\oint_{L}\nabla S\cdot d\mathbf{q}=nh.
\end{equation}

But this condition is arbitrary, Wallstrom argued, as there's no reason
in stochastic mechanics why the change in $S$ along $L$ should be
constrained to an integer multiple of $h$. Indeed, assuming this
condition amounts to assuming that wave functions are single-valued,
which amounts to assuming that the solution space of the Nelson-Yasue
stochastic mechanical equations is equivalent to the solution space
of the quantum mechanical Schrödinger equation. Such an assumption
cannot be made, however, in a theory purporting to \textit{\textcolor{black}{derive}}
the Schrödinger equation of quantum mechanics.

These arguments notwithstanding, one might question whether the requirement
of single-valued wave functions in quantum mechanics is any less arbitrary
than imposing (27) in stochastic mechanics. This is not the case.
The single-valuedness condition, as usually motivated, is a consequence
of imposing two completely natural boundary conditions on solutions
of (23): (a) that the solutions satisfy the linear superposition principle
\cite{Schroedinger1938,Wallstrom1989}, and (b) that $|\psi|^{2}$
can be physically interpreted as a probability density \cite{Bohm1951,Merzbacher1962,Matthews1979}.
\footnote{Henneberger et al. \cite{Henneberger1994} argue that the single-valuedness
condition on wave functions is strictly a consequence of the linear
superposition principle. However, this nuance is inessential to our
arguments. } Condition (a) is natural to the single-valuedness requirement because
of the linearity of the Schrödinger equation, and condition (b) is
natural to it because a probability density is, by definition, a single-valued
function on its sample space. Moreover, it can be shown that if (a)
doesn't hold then (b) doesn't hold for any linear superposition of
two or more solutions. To illustrate this, consider the free particle
Schrödinger equation on the unit circle, $\textrm{S}^{1}$: \footnote{This argument was relayed to the author by T. Wallstrom {[}private
communication{]}.} 
\begin{equation}
-\frac{\hbar^{2}}{2m}\frac{1}{r^{2}}\frac{\partial^{2}\psi}{\partial\theta^{2}}=E\psi.
\end{equation}
The un-normalized wave function satisfying this equation is of the
form $\psi(\theta)=Ne^{ik\theta}$, where $k=\frac{r}{\hbar}\sqrt{2mE}$.
For this wave function to satisfy (b), $k$ (and hence the energy
$E$) can take any positive value among the real numbers since obviously
$|\psi|^{2}=N^{2}$. Consider now a superposition of the form $\psi_{s}(\theta)=N\left(e^{ik_{1}\theta}+e^{ik_{2}\theta}\right)$,
which leads to the density 
\begin{equation}
|\psi_{s}|^{2}=2N^{2}\left(1+cos\left[(k_{1}-k_{2})\theta\right]\right).
\end{equation}

If $k_{1}$ and $k_{2}$ are allowed to take non-integer values, then
$(k_{1}-k_{2})$ can also take non-integer values, and the density
formed from the superposition can be multi-valued, thereby violating
(b). Condition (a) will also be violated since, although a single
wave function in the superposition satisfies (b), the superposition
does not; so the set of wave functions of the form $\psi(\theta)=Ne^{ik\theta}$,
where $k$ can take non-integer values, does not form a linear space.
If, however, $k_{1}$ and $k_{2}$ are integers, then so is $(k_{1}-k_{2})$,
and conditions (a) and (b) will be satisfied since $|\psi_{s}|^{2}$
will always be single-valued. Correspondingly, it follows that the
energy and momentum of the particle on the unit circle will be quantized
with $e^{i2\pi\frac{r}{\hbar}\sqrt{2mE}}=1=e^{i2\pi n}$ yielding
$E_{n}=\frac{p_{\theta}^{2}}{2mr^{2}}=\frac{n^{2}\hbar^{2}}{2mr^{2}}$,
where $n$ is an integer.

The wave functions constructed from stochastic mechanics will therefore
satisfy only (b) if $S$ is arbitrarily multi-valued, while they will
satisfy (a) and (b) together only when (27) is imposed. But as previously
mentioned, (27) is ad hoc in stochastic mechanics, and assuming it
to obtain only single-valued wave functions is logically circular
if the objective of stochastic mechanics is to derive quantum mechanics.
The challenge then is to find a physically plausible justification
for (27) strictly within the assumptions of existing formulations
of stochastic mechanics, or otherwise some new formulation. Accordingly,
we shall now begin the development of our proposed justification through
a reformulation of Nelson-Yasue stochastic mechanics (NYSM).

\section{Classical Model of Constrained Zitterbewegung Motion}

Here we develop a classical model of a particle of mass \emph{m} constrained
in its rest frame to undergo a simple harmonic oscillation of (electron)
Compton frequency, and show that it gives rise to a quantization condition
equivalent to (27). Our model motivates the quantization condition
from essentially the same physical arguments used by de Broglie in
his ``phase-wave'' model \cite{Broglie1925,Darrigol1994} and by
Bohm in his subquantum field-theoretic models \cite{Bohm1957,Bohm2002}.
However, it differs from both de Broglie's model and Bohm's models
in that we do not need to refer to fictitious ``phase waves'', nor
assume that our particle is some localized distribution of a (hypothetical)
fluctuating subquantum field \cite{Bohm1957}, nor assume a non-denumerable
infinity of ``local clocks'' at each point in space-time \cite{Bohm2002}.
We start by developing the free particle case, extend it to a classical
Hamilton-Jacobi (HJ) statistical mechanical description, and repeat
these steps with the inclusion of interactions with external fields.

The purpose of this section is three-fold: (i) to explicitly show,
without the added conceptual complications of stochastic mechanics,
the basic physical assumptions underlying our particle model; (ii)
to show how our model can be consistently generalized to include interactions
with external fields; (iii) to show, using a well-established formulation
of classical statistical mechanics that has conceptual and mathematical
similarities to stochastic mechanics, how our model can be consistently
generalized to a statistical ensemble description (which will also
be necessary in the stochastic mechanical case), and how doing so
gives a quantization condition equivalent to (27) for a `classical'
wave function satisfying a nonlinear Schrödinger equation. No attempt
will be made here to suggest a physical/dynamical model for the zitterbewegung
motion. A framework for a physical model is given in section 5, while
a discussion of possible physical models is reserved for Part II.

\subsection{One free particle}

Suppose that a classical particle of rest mass $m$ is rheonomically
constrained to undergo a periodic process with constant angular frequency,
$\omega_{0}$, about some fixed point in 3-space, $\mathbf{q}_{0}$,
in a Lorentz frame where the particle has translational velocity $\mathbf{v}=d\mathbf{q}_{0}/dt=0$.
The exact nature of this process is not important for the argument
that follows, as long as it is periodic. For example, this process
could be an oscillation or (if the particle is spinning) a rotation.
But since we are considering the spinless case, we will take the periodic
process to be some kind of oscillation. The constancy of $\omega_{0}$
implies that the oscillation is simply harmonic with phase $\theta=\omega_{0}t_{0}+\phi$.
Although the assumption of simple harmonic motion implies that $\theta$
is a continuous function of the particle's position, in the translational
rest frame, it must be the case that the phase change $\delta\theta$
at any fixed instant $t_{0}$ will be zero for some translational
displacement $\delta\mathbf{q}_{0}$. Otherwise, such a displacement
would define a preferred direction in space given by $\nabla\theta(\mathbf{q}_{0})$.
Hence, in the translational rest frame, we can write 
\begin{equation}
\delta\theta=\omega_{0}\delta t_{0},
\end{equation}
where $\delta t_{0}$ is the change in proper time.

If we Lorentz transform to the lab frame where the particle has constant
translational velocity, $\mathbf{v}$, and undergoes a displacement
$\delta\mathbf{q}(t)$ in $\delta t$, then $\delta t_{0}=\gamma\left(\delta t-\mathbf{v}\cdot\delta\mathbf{q}(t)/c^{2}\right)$
and (30) becomes 
\begin{equation}
\delta\theta(\mathbf{q}(t),t)=\omega_{0}\gamma\left(\delta t-\frac{\mathbf{v}\cdot\delta\mathbf{q}(t)}{c^{2}}\right),
\end{equation}
where $\gamma=1/\sqrt{\left(1-\mathbf{v}^{2}/c^{2}\right)}$. Recalling
that for a relativistic free particle we have $E=\gamma mc^{2}$ and
$\mathbf{p}=\gamma m\mathbf{v}$, (31) can be equivalently expressed
as 
\begin{equation}
\delta\theta(\mathbf{q}(t),t)=\frac{\omega_{0}}{mc^{2}}\left(E\delta t-\mathbf{p}\cdot\delta\mathbf{q}(t)\right).
\end{equation}
Suppose now that the oscillating particle is physically or virtually
\footnote{Because we permit a virtual displacement where time changes, we cannot
use the definition of a virtual displacement often found in textbooks
\cite{H.Goldstein2011,JoseSaletan} (which assumes time is fixed under
the displacement). Instead, we use the more refined definition of
virtual displacements proposed by Ray \& Shamanna \cite{Ray2006},
namely that a virtual displacement is the difference between any two
(unequal) ``allowed displacements'', or $\delta\mathbf{q}_{k}=d\mathbf{q}_{k}-d\mathbf{q}_{k}^{'}$,
where $k=1,2,...,N,$ and an allowed displacement is defined as $d\mathbf{q}_{k}=\mathbf{v}_{k}dt,$
where $\mathbf{v}_{k}$ are the ``virtual velocities'', or the velocities
allowed by the mechanical constraints of a given system.} displaced around a closed loop $L$ (i.e., a continuous, non-self-intersecting
loop that is otherwise arbitrary) in which both position and time
can vary. The consistency of the model requires that the accumulated
phase change be given by 
\begin{equation}
\oint_{L}\delta\theta(\mathbf{q}(t),t)=\frac{\omega_{0}}{mc^{2}}\oint_{L}\left(E\delta t-\mathbf{p}\cdot\delta\mathbf{q}(t)\right)=2\pi n,
\end{equation}
where $n$ is an integer. This follows from the assumption that the
oscillation is simply harmonic in the particle's rest frame, which
makes $\theta$ in the lab frame a single-valued function of $\mathbf{q}(t)$
(up to an additive integer multiple of $2\pi$). Indeed, if (33) were
not true, we would contradict our hypothesis that the oscillating
particle has a well-defined phase at each point along its space-time
trajectory.

If we further make the `zitterbewegung' (\emph{zbw}) hypothesis that
$m=m_{e}=9.11\times10^{-28}g$ and $\omega_{0}/m_{e}c^{2}=1/\hbar$
so that $\omega_{0}=\omega_{c}=7.77\times10^{20}rad/s,$ which is
the electron Compton angular frequency, then we can define $\bar{\theta}\eqqcolon-\frac{1}{\hbar}S$
and (33) can be rewritten as 
\begin{equation}
\oint_{L}\delta S(\mathbf{q}(t),t)=\oint_{L}\left(\mathbf{p}\cdot\delta\mathbf{q}(t)-E\delta t\right)=nh.
\end{equation}
Finally, for the special case of loop integrals in which time is held
fixed ($\delta t=0$), (34) reduces to 
\begin{equation}
\oint_{L}\mathbf{p}\cdot\delta\mathbf{q}(t)=nh,
\end{equation}
which we may observe is formally identical to the Bohr-Sommerfeld-Wilson
quantization condition.

By integrating (32) and using the Legendre transformation, it can
be shown that the phase of the free \emph{zbw} particle is, equivalently,
its relativistic action up to an additive constant, or $S(\mathbf{q}(t),t)=\mathbf{p}\cdot\mathbf{q}(t)-Et-\hbar\phi=-mc^{2}\int_{t_{i}}^{t}dt'/\gamma+C$.
\footnote{The proof is as follows. From $L=-mc^{2}/\gamma$, the Legendre transform
gives $E=\mathbf{p}\cdot\mathbf{v}-L=\gamma mv^{2}+mc^{2}/\gamma=\gamma mc^{2}$
and $L=\mathbf{p}\cdot\mathbf{v}-E.$ So for the free \emph{zbw} particle,
$S=\int Ldt+C=\int\left(\mathbf{p}\cdot\mathbf{v}-E\right)dt+C=\int\left(\mathbf{p}\cdot d\mathbf{q}-Edt\right)+C=\mathbf{p}\mathbf{\cdot q}-Et+C$
(absorbing the integration constants arising from $d\mathbf{q}$ and
$dt$ into \emph{C}).}, where $\phi$ is the initial phase constant. Recognizing also that
$\mathbf{p}=\hbar\gamma\omega_{c}\mathbf{v}/c^{2}=\hbar\gamma\mathbf{k}$
and $E=\hbar\gamma\omega_{c}$, the translational 3-velocity of the
particle can be obtained from $S(\mathbf{q}(t),t)$ as $\mathbf{v}=(1/\gamma m)\nabla S(\mathbf{q},t)|_{\mathbf{q}=\mathbf{q}(t)}$,
and the total relativistic energy as $E=-\partial_{t}S(\mathbf{q},t)|_{\mathbf{q}=\mathbf{q}(t)}$.
It follows then that $S(\mathbf{q}(t),t)$ is a solution of the classical
relativistic Hamilton-Jacobi equation, 
\begin{equation}
-\partial_{t}S(\mathbf{q},t)|_{\mathbf{q}=\mathbf{q}(t)}=\sqrt{m^{2}c^{4}+\left(\nabla S(\mathbf{q},t)\right)^{2}c^{2}}|_{\mathbf{q}=\mathbf{q}(t)}.
\end{equation}
In the non-relativistic limit, $v\ll c$, $S(\mathbf{q}(t),t)\approx m\mathbf{v}\cdot\mathbf{q}(t)-\left(mc^{2}+\frac{mv^{2}}{2}\right)t-\hbar\phi$,
and (36) becomes 
\begin{equation}
-\partial_{t}S(\mathbf{q},t)|_{\mathbf{q}=\mathbf{q}(t)}=\frac{\left(\nabla S(\mathbf{q},t)\right)^{2}}{2m}|_{\mathbf{q}=\mathbf{q}(t)}+mc^{2},
\end{equation}
where $\mathbf{v}=\left(1/m\right)\nabla S|_{\mathbf{q}=\mathbf{q}(t)}=(1/m)\hbar\mathbf{k}$
and satisfies the trivial classical Newtonian equation 
\begin{equation}
m\mathbf{a}=\left(\frac{\partial}{\partial t}+\mathbf{v}\cdot\nabla\right)\nabla S=0.
\end{equation}
We find then that, in the non-relativistic limit, the oscillation
frequency of the \emph{zbw} particle has two parts - a low frequency
oscillation, $\omega_{dB}=\hbar k^{2}/2m$, which modulates the high
frequency oscillation $\omega_{c}$.

Evidently (37) has the form of the non-relativistic dispersion relation
$E=\hbar^{2}k^{2}/2m+mc^{2}$, which naively suggests that one can
obtain the free-particle Schrödinger equation for a plane wave by
introducing operators $\hat{p}=-i\hbar\nabla$ and $\hat{E}=i\hbar\partial_{t}$
such that $\hat{p}\psi=\hbar k\psi$, $\hat{E}\psi=\hbar\omega\psi$,
and $i\hbar\partial_{t}\psi=-\left(\hbar^{2}/2m\right)\nabla^{2}\psi$
for $\psi(\mathbf{q},t)=Ae^{i\left(\mathbf{p}\cdot\mathbf{q}-Et\right)/\hbar}.$
However, there is no physical wave for such a plane wave to be identified
with in our model. Such a plane wave and Schrödinger equation are
nothing more than abstract, mathematically equivalent re-writings
of the \emph{zbw} particle energy equation (37). On the other hand,
as we will see next, a \textit{\textcolor{black}{nonlinear}} Schrödinger
equation that describes the dynamical evolution of a statistical ensemble
of identical \emph{zbw} particles is derivable from the classical
HJ description of the ensemble.

\subsection{Classical Hamilton-Jacobi statistical mechanics for one free particle}

Suppose that the actual position and momentum of a \emph{zbw} particle,
$\left(\mathbf{q}(t),\mathbf{p}(t)\right)$, are unknown. Then we
must resort to the description of a classical (i.e., Gibbsian) statistical
ensemble of fictitious, identical, non-interacting \emph{zbw} particles
\cite{Holland1993}, which differ from each other only by virtue of
their initial positions, velocities, and (possibly) phases. (Consideration
of this in the classical context will be helpful for seeing how our
model can be incorporated into stochastic mechanics.) In terms of
the \emph{zbw} phase, this change in description corresponds to replacing
$\delta S(\mathbf{q}(t),t)$ by $dS(\mathbf{q},t)=\mathbf{p}(\mathbf{q},t)\cdot\mathbf{\mathit{d}q}-E(\mathbf{q},t)dt$,
which we obtained from replacing $\mathbf{q}(t)$ by $\mathbf{q}$,
where $\mathbf{q}$ labels a \textit{\textcolor{black}{possible}}
position in 3-D space that the actual \emph{zbw} particle could occupy
at time $t$. Integrating $dS(\mathbf{q},t)$ then gives $S(\mathbf{q},t)=\int\mathbf{p}(\mathbf{q},t)d\mathbf{q}-\int E(\mathbf{q},t)dt+C$,
where $C=\hbar\phi$ is just the initial phase constant. So $S(\mathbf{q},t)$
is a phase \textit{\textcolor{black}{field}} connected with the ensemble,
$\mathbf{p}(\mathbf{q},t)=\nabla S(\mathbf{q},t)$ is the corresponding
translational momentum field, and $E(\mathbf{q},t)=-\partial_{t}S(\mathbf{q},t)$
is the total energy field. Note that, for any initial $\mathbf{q}$
and $t$, the constant $\phi$ can be given any value on the interval
$\left[0,2\pi\right]$; i.e., the initial phase constant associated
with any member of the ensemble can be freely specified on that interval.
(Of course, this phase constant does not affect the momentum field
or the total energy field, as these fields are obtained from space-time
derivatives of the phase field. Thus there are many phase fields corresponding
to a unique momentum field and total energy field.)

Now, in the specific case of the free \emph{zbw} particle, $p=const$
and $E=const$ for each member of the ensemble. So the infinitesimal
phase change connected with the ensemble is just $dS(\mathbf{q},t)=\mathbf{p}\cdot\mathbf{\mathit{d}q}-Edt$,
yielding $S(\mathbf{q},t)=\mathbf{p}\cdot\mathbf{q}-Et+C$ upon integration.

With this phase field in hand, we can now construct a classical HJ
statistical mechanics for our \emph{zbw} particle. Essentially, $S(\mathbf{q},t)$
and $\nabla S(\mathbf{q},t)$ will respectively satisfy the classical
Hamilton-Jacobi equation, 
\begin{equation}
-\partial_{t}S(\mathbf{q},t)=\frac{\left(\nabla S(\mathbf{q},t)\right)^{2}}{2m}+mc^{2},
\end{equation}
and the trivial classical Newtonian equation, 
\begin{equation}
m\mathbf{a}(\mathbf{q},t)=\left(\frac{\partial}{\partial t}+\mathbf{v}(\mathbf{q},t)\cdot\nabla\right)\nabla S(\mathbf{q},t)=0.
\end{equation}
If we now suppose that the density of ensemble particles per unit
volume in an element $d^{3}q$ surrounding the point $\mathbf{q}$
at time $t$ is given by the function $\rho(\mathbf{q},t)\geq0$,
which satisfies the normalization condition $\int\rho_{0}(\mathbf{q})d^{3}q=1$,
then it is straightforward to show \cite{Holland1993} that $\rho(\mathbf{q},t)$
evolves in time by the continuity equation 
\begin{equation}
\frac{\partial\rho({\normalcolor \mathbf{q},t})}{\partial t}=-\nabla\cdot\left[\mathbf{\frac{\nabla\mathrm{\mathit{S\mathrm{(\mathbf{q},\mathit{t})}}}}{\mathit{m}}}\rho(\mathbf{q},t)\right].
\end{equation}
Accordingly, $\rho(\mathbf{q},t)$ carries the interpretation of the
probability density for the actual \emph{zbw} particle position $\mathbf{q}(t)$.
And since $S(\mathbf{q},t)$ is a field over the possible positions
that the actual \emph{zbw} particle can occupy at time \emph{t}, where
for each possible position the actual \emph{zbw} particle's phase
will satisfy the relation (35), $S(\mathbf{q},t)$ will be a single-valued
function of $\mathbf{q}$ and $t$ (up to an additive integer multiple
of $2\pi$) and satisfy 
\begin{equation}
\oint_{L}dS(\mathbf{q},t)=\oint_{L}\nabla S(\mathbf{q},t)\cdot d\mathbf{q}=nh.
\end{equation}
The use of exact differentials in (42) indicates that the loop integral
is now an integral of the momentum field along any closed \textit{mathematical}
loop in 3-space with time held constant; that is, a closed loop around
which the actual particle with momentum $\mathbf{p}$ \emph{could
potentially be displaced}, starting from any possible position $\mathbf{q}$
it can occupy at fixed time $t$. This tells us that the circulation
of the momentum field is quantized, in contrast to an ordinary classical
statistical mechanical ensemble for which the momentum field circulation
need not satisfy (42).

Finally, we can combine (39) and (41) into the nonlinear Schrödinger
equation \cite{Schiller1962,Rosen1964,Holland1993,Ghose2002,Nikolic2006,Nikolic2007},
\begin{equation}
i\hbar\frac{\partial\psi(\mathbf{q},t)}{\partial t}=-\frac{\hbar^{2}}{2m}\nabla^{2}\psi(\mathbf{q},t)+\frac{\hbar^{2}}{2m}\frac{\nabla^{2}|\psi(\mathbf{q},t)|}{|\psi(\mathbf{q},t)|}\psi(\mathbf{q},t)+mc^{2}\psi(\mathbf{q},t),
\end{equation}
with general solution $\psi(\mathbf{q},t)=\sqrt{\rho_{0}(\mathbf{q}-\mathbf{v}_{0}t)}e^{iS(\mathbf{q},t)/\hbar}$,
which is single-valued because of (42). (Note that $C$ will contribute
a global phase factor, $e^{iC/\hbar}$, which cancels out from both
sides.) As an example of a specific solution, the complex phase $e^{iS/\hbar}$
takes the form of a plane-wave, $S=\mathbf{p}\cdot\mathbf{q}-Et+\hbar\phi$,
while the initial probability density, $\rho_{0}$, can take the form
of a Gaussian that propagates with fixed profile and speed $v$ (in
contrast to a Gaussian density in free particle quantum mechanics,
which disperses over time).

We have thereby shown that extending our free \emph{zbw} particle
model to a classical HJ statistical mechanics allows us to derive
a nonlinear Schrödinger equation with single-valued wave functions.
Next we will incorporate interactions of the \emph{zbw} particle with
external fields.

\subsection{One particle interacting with external fields}

To describe the interaction of our \emph{zbw} particle with fields,
let us reconsider the change in the \emph{zbw} phase in the rest frame.
In terms of the rest energy of the \emph{zbw} particle, we can rewrite
(30) as 
\begin{equation}
\delta\theta=\omega_{c}\delta t_{0}=\frac{1}{\hbar}\left(mc^{2}\right)\delta t_{0}.
\end{equation}
Any additional contribution to the energy of the particle, such as
from a weak external gravitational field (e.g. the Earth's gravitational
field) coupling to the particle's mass $m$ via $\Phi_{g}=\mathbf{g\cdot q}$,
will then modify (44) as 
\begin{equation}
\delta\theta=\left(\omega_{c}+\kappa(\mathbf{q})\right)|_{\mathbf{q}=\mathbf{q}_{0}}\delta t_{0}=\frac{1}{\hbar}\left(mc^{2}+m\Phi_{g}(\mathbf{q})\right)|_{\mathbf{q}=\mathbf{q}_{0}}\delta t_{0},
\end{equation}
where $\kappa=\omega_{c}\Phi_{g}/c^{2}$. In other words, the gravitational
field shifts the \emph{zbw} frequency in the rest frame by a very
small amount. For example, if $|\mathbf{g}|=\mathrm{10^{3}\mathit{cm/s^{2}}}$
and is in the $\mathbf{\hat{z}}$ direction, and we take $|\mathbf{q}|=100cm$,
then $\kappa\approx\omega_{c}\times10^{-16}.$ Here we have approximated
the point at which the \emph{zbw} particle interacts with the external
gravitational field to be just its equilibrium position, $\mathbf{q}_{0}$,
because the displacement $|\mathbf{q}|\gg\lambda_{c},$ allowing us
to approximate the interaction with the mass as point-like. \footnote{This appears to be the same assumption made by de Broglie for his
equivalent model, although he never explicitly says so. Bohm, to the
best of our knowledge, never extended his models to include field
interactions.}

In addition, we could allow the \emph{zbw} particle to carry charge
$e$ (so that it now becomes a classical charged oscillator, subject
to the hypothetical constraint that it does not radiate electromagnetic
energy in its rest frame, or the constraint that the oscillation of
the charge is radially symmetric so that there is no net energy radiated
\cite{Schott1,Schott2,Schott3}, or constrained to correspond to one
of the non-spherically-symmetric charge distributions considered by
Bohm and Weinstein \cite{BohmWeinstein1948} for which the retarded
self-fields cause the charge distribution to oscillate at a fixed
frequency without radiating) which couples to an external (and possibly
space-time varying) electric field such that $\Phi_{e}=\mathbf{E(q},t)\cdot\mathbf{q}$,
where \textbf{$\mathbf{q}$} is the displacement vector in some arbitrary
direction from the field source. Here again we can make the point-like
approximation, as in laboratory experiments the displacement of a
particle from a field source is typically on the centimeter scale,
making $|\mathbf{q}|\gg\lambda_{c}$). Then 
\begin{equation}
\delta\theta=\left(\omega_{c}+\kappa(\mathbf{q}_{0})+\varepsilon(\mathbf{q}_{0},t_{0})\right)\delta t_{0}=\frac{1}{\hbar}\left(mc^{2}+m\Phi_{g}(\mathbf{q}_{0})+e\Phi_{e}(\mathbf{q}_{0},t_{0})\right)\delta t_{0},
\end{equation}
where $\varepsilon=\omega_{c}\left(e/mc^{2}\right)\Phi_{e}$. Assuming
$\mathbf{E}$ has an experimental value of $\sim10^{5}V/cm\approx.03stV/cm$,
which is the upper limit laboratory field strength that can be produced
in Stark effect experiments \cite{Bichsel2007}, and $|\mathbf{q}|=1cm$,
then $\varepsilon\approx\omega_{c}\times10^{-5}$, which is also a
very small shift.

If we now transform to the laboratory frame where the \emph{zbw} particle
has nonzero but variable translational velocity, (46) becomes 
\begin{equation}
\begin{aligned}\delta\theta(\mathbf{q}(t),t) & =\left[\left(\omega_{dB}+\kappa(\mathbf{q})+\varepsilon(\mathbf{q})\right)\gamma\left(\delta t-\frac{\mathbf{v}_{0}(\mathbf{q},t)\cdot\delta\mathbf{q}}{c^{2}}\right)\right]_{\mathbf{q}=\mathbf{q}(t)}\\
 & =\frac{1}{\hbar}\left[\left(\gamma mc^{2}+\gamma m\Phi_{g}(\mathbf{q})+e\Phi_{e}(\mathbf{q},t)\right)\delta t\right.\\
 & \left.-\left(\gamma mc^{2}+\gamma m\Phi_{g}(\mathbf{q})+e\Phi_{e}(\mathbf{q},t)\right)\frac{\mathbf{v}_{0}(\mathbf{q},t)\cdot\delta\mathbf{q}}{c^{2}}\right]|_{\mathbf{q}=\mathbf{q}(t)}\\
 & =\frac{1}{\hbar}\left(E(\mathbf{q}(t),t)\delta t-\mathbf{p}(\mathbf{q}(t),t)\cdot\delta\mathbf{q}(t)\right),
\end{aligned}
\end{equation}
where $E=\gamma mc^{2}+\gamma m\Phi_{g}+e\Phi_{e}$ and $\mathbf{p}=m\mathbf{v}=\left(\gamma mc^{2}+\gamma m\Phi_{g}+e\Phi_{e}\right)\left(\mathbf{v}_{0}/c^{2}\right)$.
(Note that the term $e\Phi_{e}$ is unaffected by the Lorentz transformation
because it doesn't involve the particle's rest mass.) Here the velocity
$\mathbf{v}_{0}$ is that of a free particle, while $\mathbf{v}$
is the adjusted velocity due to the presence of external potentials.
In this moving frame, we can also have the \emph{zbw} particle couple
to an external magnetic vector potential \footnote{We could of course also include a gravitational vector potential,
but for simplicity we'll just stick with the magnetic version. } such that $\mathbf{v}\rightarrow\mathbf{v}'=\mathbf{v}+e\mathbf{A}_{ext}/\gamma mc$
(and $\gamma$ depends on $v$). Although the physical influence of
the fields now allows the $\omega$ and $\mathbf{k}$ of the particle
to vary as a function of position and time, the phase of the oscillation
is still a well-defined function of the particle's space-time location;
so if we displace the oscillating particle around a closed loop, the
phase change is still given by 
\begin{equation}
\oint_{L}\delta\theta(\mathbf{q}(t),t)=\frac{1}{\hbar}\oint_{L}\left(E(\mathbf{q}(t),t)\delta t-\mathbf{p}'(\mathbf{q}(t),t)\cdot\delta\mathbf{q}(t)\right)=2\pi n,
\end{equation}
or 
\begin{equation}
\oint_{L}\delta S(\mathbf{q}(t),t)=\oint_{L}\left(\mathbf{p}'(\mathbf{q}(t),t)\cdot\delta\mathbf{q}(t)-E(\mathbf{q}(t),t)\delta t\right)=nh.
\end{equation}
For the special case of a loop in which time is held fixed, we then
have 
\begin{equation}
\oint_{L}\nabla S(\mathbf{\mathbf{q}},t)|_{\mathbf{q}=\mathbf{q}(t)}\cdot\delta\mathbf{q}(t)=\oint_{L}\mathbf{p}'(\mathbf{q}(t),t)\cdot\delta\mathbf{q}(t)=nh,
\end{equation}
or 
\begin{equation}
\oint_{L}m\mathbf{v}(\mathbf{\mathbf{q}}(t),t)\cdot\delta\mathbf{q}(t)=nh-\frac{e}{c}\oint_{L}\mathbf{A}_{ext}(\mathbf{q}(t),t)\cdot\delta\mathbf{q}(t),
\end{equation}
where the last term on the right hand side of (51) is, by Stokes'
theorem, the magnetic flux enclosed by the loop.

We can also integrate (47) and rewrite in terms of $S(\mathbf{q}(t),t)$
to obtain 
\begin{equation}
S(\mathbf{q}(t),t)=\int_{\mathbf{q}_{i}(t_{i})}^{\mathbf{q}(t)}\mathbf{p}'(\mathbf{q}(s),s)\cdot d\mathbf{q}(s)-\int_{t_{i}}^{t}E(\mathbf{q}(s),s)ds-\hbar\phi,
\end{equation}
where $\phi$ is the initial phase constant and (52) is equivalent
(up to an additive constant) to the relativistic action of a particle
in the presence of external fields. \footnote{The proof is as follows. From $L=-mc^{2}/\gamma-\gamma m\Phi_{g}-e\Phi_{e}+e\mathbf{\frac{v}{c}}\cdot\mathbf{A}_{ext}$,
the Legendre transform gives $E=\mathbf{p}'\cdot\mathbf{v}-L=\gamma mv{}^{2}+mc^{2}/\gamma+\gamma m\Phi_{g}+e\Phi_{e}=\gamma mc^{2}+\gamma m\Phi_{g}+e\Phi_{e}$
and $L=\mathbf{p}'\cdot\mathbf{v}-E$. So, $S=\int Ldt+C=\int\left(\mathbf{p}'\cdot\mathbf{v}-E\right)dt+C=\int\left(\mathbf{p}'\cdot d\mathbf{q}-Edt\right)+C$.} As before, the translational kinetic 3-velocity of the particle can
be obtained from $S(\mathbf{q}(t),t)$ as $\mathbf{v}(\mathbf{q}(t),t)=\mathbf{p}(\mathbf{q}(t),t)/\gamma m=(1/\gamma m)\nabla S(\mathbf{q},t)|_{\mathbf{q}=\mathbf{q}(t)}-e\mathbf{A}_{ext}(\mathbf{q}(t),t)/\gamma mc$,
and the total relativistic energy as $E(\mathbf{q}(t),t)=-\partial_{t}S(\mathbf{q},t)|_{\mathbf{q}=\mathbf{q}(t)}$.
It then follows that $S(\mathbf{q}(t),t)$ is a solution of the classical
relativistic Hamilton-Jacobi equation 
\begin{equation}
-\partial_{t}S(\mathbf{q},t)|_{\mathbf{q}=\mathbf{q}(t)}=\sqrt{m^{2}c^{4}+\left(\nabla S(\mathbf{q},t)-\frac{e}{c}\mathbf{A}_{ext}(\mathbf{q},t)\right)^{2}c^{2}}|_{\mathbf{q}=\mathbf{q}(t)}+\gamma m\Phi_{g}(\mathbf{q}(t))+e\Phi_{e}(\mathbf{q}(t),t).
\end{equation}
When $v\ll c,$ 
\begin{equation}
\begin{aligned}S(\mathbf{q}(t),t) & \approx\int_{\mathbf{q}_{i}(t_{i})}^{\mathbf{q}(t)}m\mathbf{v'(q}(s),s)\cdot d\mathbf{q}(s)-\\
 & -\int_{t_{i}}^{t}\left(mc^{2}+\frac{1}{2m}\left[\mathbf{p}(\mathbf{q}(s),s)-\frac{e}{c}\mathbf{A}_{ext}(\mathbf{q}(s),s)\right]^{2}+m\Phi_{g}(\mathbf{q}(s))+e\Phi_{e}(\mathbf{q}(s),s)\right)ds-\hbar\phi,
\end{aligned}
\end{equation}
and (53) becomes 
\begin{equation}
-\partial_{t}S(\mathbf{q},t)|_{\mathbf{q}=\mathbf{q}(t)}=\frac{\left(\nabla S(\mathbf{q},t)-\frac{e}{c}\mathbf{A}_{ext}(\mathbf{q},t)\right)^{2}}{2m}|_{\mathbf{q}=\mathbf{q}(t)}+mc^{2}+m\Phi_{g}(\mathbf{q}(t))+e\Phi_{e}(\mathbf{q}(t),t),
\end{equation}
with $\mathbf{v}(\mathbf{q}(t),t)=(1/m)\nabla S(\mathbf{q},t)|_{\mathbf{q}=\mathbf{q}(t)}-e\mathbf{A}_{ext}(\mathbf{q}(t),t)/mc$
and satisfies the classical Newtonian equation of motion, 
\begin{equation}
\begin{aligned}m\mathbf{a}(\mathbf{q}(t),t) & =\left(\frac{\partial}{\partial t}+\mathbf{v}(\mathbf{q}(t),t)\cdot\nabla\right)\left[\nabla S(\mathbf{q},t)-\frac{e}{c}\mathbf{A}_{ext}(\mathbf{q},t)\right]|_{\mathbf{q}=\mathbf{q}(t)}\\
 & =-\nabla\left[m\Phi_{g}(\mathbf{q})+e\Phi_{e}(\mathbf{q},t)\right]|_{\mathbf{q}=\mathbf{q}(t)}-\frac{e}{c}\frac{\partial\mathbf{A}_{ext}(\mathbf{q},t)}{\partial t}|_{\mathbf{q}=\mathbf{q}(t)}+\frac{e}{c}\mathbf{v(q}(t),t)\times\mathbf{B}_{ext}(\mathbf{q}(t),t).
\end{aligned}
\end{equation}

Incidentally, if we choose $\Phi_{e}$ as the Coulomb potential for
the hydrogen atom and set $\mathbf{B}_{ext}=0$, then our model is
empirically equivalent to the Bohr model of the hydrogen atom (the
demonstration of this can be found in Appendix B). As in the previous
section, we now want to extend our model to a classical HJ statistical
mechanics.

\subsection{Classical Hamilton-Jacobi statistical mechanics for one particle
interacting with external fields}

Suppose now that, in the lab frame with $v\ll c$, we do not know
the actual position $\mathbf{q}(t)$ of the \emph{zbw} particle. Then
the phase (54) becomes the phase field 
\begin{equation}
\begin{aligned}S(\mathbf{q},t) & =\int_{\mathbf{q}(t_{i})}^{\mathbf{q}(t)}m\mathbf{v'(q}(s),s)\cdot\mathbf{\mathit{d}q}(s)|_{\mathbf{q}(t)=\mathbf{q}}\\
 & -\int_{t_{i}}^{t}\left(mc^{2}+\frac{1}{2m}\left[\mathbf{p}(\mathbf{q}(s),s)-\frac{e}{c}\mathbf{A}_{ext}(\mathbf{q}(s),s)\right]^{2}+m\Phi_{g}(\mathbf{q}(s))+e\Phi_{c}(\mathbf{q}(s),s)\right)ds|_{\mathbf{q}(t)=\mathbf{q}}-\hbar\phi.
\end{aligned}
\end{equation}
To obtain the equations of motion for $\ensuremath{S(\mathbf{q},t)}$
and $\ensuremath{\mathbf{v}(\mathbf{q},t)}$ we will apply the classical
analogue of Yasue's variational principle, in anticipation of the
method we will use for constructing ZSM.

First we introduce the ensemble-averaged action/phase functional (inputting
limits between initial and final states), 
\begin{equation}
\begin{aligned}J & =\mathrm{E}\left[\int_{\mathbf{q}(t_{i})}^{\mathbf{q}(t_{f})}m\mathbf{v'}\cdot\mathbf{\mathit{d}q}(t)-\int_{t_{i}}^{t_{f}}\left(mc^{2}+\frac{1}{2m}\left[\mathbf{p}-\frac{e}{c}\mathbf{A}_{ext}\right]^{2}+m\Phi_{g}+e\Phi_{e}\right)dt-\hbar\phi\right]\\
 & =\mathrm{E}\left[\int_{t_{i}}^{t_{f}}\left\{ \frac{1}{2}m\mathbf{v}^{2}+\frac{e}{c}\mathbf{A}_{ext}\cdot\mathbf{v}-mc^{2}-m\Phi_{g}-e\Phi_{e}\right\} dt-\hbar\phi\right],
\end{aligned}
\end{equation}
where the equated expressions are related by the usual Legendre transformation.
Imposing the variational constraint, 
\begin{equation}
J=extremal,
\end{equation}
a straightfoward computation exactly along the lines of that in Appendix
A yields (56), which, upon replacing $\mathbf{q}(t)$ by $\mathbf{q}$,
corresponds to the classical Newtonian equation, 
\begin{equation}
\begin{aligned}m\mathbf{a}(\mathbf{q},t) & =\left(\frac{\partial}{\partial t}+\mathbf{v}(\mathbf{q},t)\cdot\nabla\right)\left[\nabla S(\mathbf{q},t)-\frac{e}{c}\mathbf{A}_{ext}(\mathbf{q},t)\right]\\
 & =-\nabla\left[m\Phi_{g}(\mathbf{q})+e\Phi_{e}(\mathbf{q},t)\right]-\frac{e}{c}\frac{\partial\mathbf{A}_{ext}(\mathbf{q},t)}{\partial t}+\frac{e}{c}\mathbf{v(q},t)\times\mathbf{B}_{ext}(\mathbf{q},t),
\end{aligned}
\end{equation}
where $\mathbf{v}(\mathbf{q},t)=(1/m)\nabla S(\mathbf{q},t)-e\mathbf{A}_{ext}(\mathbf{q},t)/mc$
corresponds to the kinetic velocity field. By integrating both sides
and setting the integration constant equal to the rest mass, we then
obtain the classical Hamilton-Jacobi equation for (57), 
\begin{equation}
-\partial_{t}S(\mathbf{q},t)=\frac{\left(\nabla S(\mathbf{q},t)-\frac{e}{c}\mathbf{A}_{ext}(\mathbf{q},t)\right)^{2}}{2m}+mc^{2}+m\Phi_{g}(\mathbf{q})+e\Phi_{e}(\mathbf{q},t).
\end{equation}
Because the momentum field couples to the vector potential, it can
be readily shown that $\rho(\mathbf{q},t)$ now evolves by the modified
continuity equation 
\begin{equation}
\frac{\partial\rho({\normalcolor \mathbf{q},t})}{\partial t}=-\nabla\cdot\left[\left(\mathbf{\frac{\nabla\mathrm{\mathit{S\mathrm{(\mathbf{q},\mathit{t})}}}}{\mathit{m}}}-\frac{e}{mc}\mathbf{A}_{ext}(\mathbf{q},t)\right)\rho(\mathbf{q},t)\right],
\end{equation}
which preserves the normalization, $\int\rho_{0}(\mathbf{q})d^{3}q=1$.
As before, $S(\mathbf{q},t)$ is a field over the possible positions
that the actual \emph{zbw} particle can occupy at time \emph{t}. Since
for each possible position the actual \emph{zbw} particle's phase
will satisfy the relation (50), $S(\mathbf{q},t)$ will be a single-valued
function of $\mathbf{q}$ and $t$ (up to an additive integer multiple
of $2\pi$) and 
\begin{equation}
\oint_{L}\nabla S(\mathbf{q},t)\cdot d\mathbf{q}=nh.
\end{equation}
Finally, we can combined (61) and (62) into the nonlinear Schrödinger
equation, 
\begin{equation}
i\hbar\frac{\partial\psi}{\partial t}=\frac{\left[-i\hbar\nabla-\frac{e}{c}\mathbf{A}_{ext}\right]^{2}}{2m}\psi+\frac{\hbar^{2}}{2m}\frac{\nabla^{2}|\psi|}{|\psi|}\psi+m\Phi_{g}\psi+e\Phi_{e}\psi+mc^{2}\psi,
\end{equation}
with wave function $\psi(\mathbf{q},t)=\sqrt{\rho(\mathbf{q},t)}e^{iS(\mathbf{q},t)/\hbar}$,
which is single-valued because of (63). (Again, $C$ will contribute
a global phase $e^{iC/\hbar}$ which drops out.)

\section{Zitterbewegung Stochastic Mechanics}

We are now ready to extend the classical \emph{zbw} model developed
in section 4 to Nelson-Yasue stochastic mechanics for all the same
cases. In doing so, we will show how this `zitterbewegung stochastic
mechanics' (ZSM) avoids the Wallstrom criticism and explain the `quantum-classical
correspondence' between the ZSM equations and the classical HJ statistical
mechanical equations. We will also apply ZSM to the central potential
problem considered by Wallstrom, to demonstrate how angular momentum
quantization emerges and therefore that the solution space of ZSM's
HJM equations is equivalent to the solution space of the quantum mechanical
Schrödinger equation.

\subsection{One free particle}

As in NYSM, we take as our starting point that a particle of rest
mass $m$ is immersed in Nelson's hypothesized ether and has a 3-space
coordinate $\mathbf{q}(t)$ undergoes a frictionless diffusion process
according to the stochastic differential equations, 
\begin{equation}
d\mathbf{q}(t)=\mathbf{b}(\mathbf{q}(t),t)dt+d\mathbf{W}(t),
\end{equation}
for the forward-time direction, and 
\begin{equation}
d\mathbf{q}(t)=\mathbf{b}_{*}(\mathbf{q}(t),t)dt+d\mathbf{W}_{*}(t),
\end{equation}
for the backward-time direction. As in NYSM, $d\mathbf{W}$ is the
Wiener process satisfying $\mathrm{E}_{t}\left[d\mathbf{W}\right]=0$
and $\mathrm{E}_{t}\left[d\mathbf{W}^{2}\right]=\left(\hbar/m\right)dt$.
Now, in order to incorporate the \emph{zbw} oscillation as a property
of the particle, we must amend Nelson's original phenomenological
hypotheses about his ether and particle with the following additional
hypotheses of phenomenological character: \footnote{Meaning, we will follow Nelson's approach of provisionally not offering
an explicit physical model of the ether, and de Broglie-Bohm's approach
of provisionally not offering an explicit physical model for the \emph{zbw}
particle, beyond the hypothetical characteristics listed here. However,
these characteristics should be regarded as general constraints on
any future physical model of Nelson's ether, the \emph{zbw} particle,
and the dynamical coupling between the two.} 
\begin{enumerate}
\item Nelson's ether is not only a stochastically fluctuating medium in
space-time, but an oscillating medium with a spectrum of angular frequencies
superposed at each point in 3-space. More precisely, we imagine the
ether as a continuous (or effectively continuous) medium composed
of a countably infinite number of fluctuating, stationary, spherical
waves \footnote{These ether waves could be fundamentally continuous field variables
or perhaps collective modes arising from nonlinear coupling between
(hypothetical) discrete constituents of the ether. Both possibilities
are logically compatible with what follows.} superposed at each point in space, with each wave having a different
(constant) angular frequency, $\omega_{0}^{k}$, where $k$ denotes
the \emph{k}-th ether mode. (If we assume an upper frequency cut-off
for our modes as the inverse Planck time, this will imply an upper
bound on the Compton frequency of an elementary particle immersed
in the ether, as we will see from hypothesis 3 below.) The relative
phases between the modes are taken to be random so that each mode
is effectively uncorrelated with every other mode. 
\item The particle of rest mass \emph{m}, located in its instantaneous mean
forward translational rest frame (IMFTRF), i.e., the frame in which
$D\mathbf{q}(t)=\mathbf{b}(\mathbf{q}(t),t)=0$, at some point $\mathbf{q}_{0}$,
is bounded to a harmonic oscillator potential with fixed natural frequency
$\omega_{0}=\omega_{c}=\left(1/\hbar\right)mc^{2}$. In keeping with
the phenomenological approach of ZSM and the approach taken by de
Broglie and Bohm with their \emph{zbw} models, we need not specify
the precise physical nature of this harmonic oscillator potential.
This is task is left for a future physical model of the ZSM particle. 
\item The particle's center of mass, as a result of being immersed in the
ether, undergoes an approximately frictionless translational Brownian
motion (due to the homogeneous and isotropic ether fluctuations that
couple to the particle by possibly electromagnetic, gravitational,
or some other means), as already described by (65-66); and, in its
IMFTRF, undergoes a driven oscillation about $\mathbf{q}_{0}$ by
coupling to a narrow band of ether modes that resonantly peaks around
the particle's natural frequency. However, in order that the oscillation
of the particle doesn't become unbounded in its kinetic energy, there
must be some mechanism by which the particle dissipates energy back
into the ether modes so that, on the average, a steady-state equilibrium
regime is reached for the oscillation. That is to say, on some hypothetical
characteristic short time-scale, $\tau$, the average energy absorbed
from the driven oscillation by the resonant ether modes equals the
average energy dissipated back to the ether by the particle. We note
that the average, in the present sense, would be over the random phases
of the ether modes. (Here we are taking inspiration from stochastic
electrodynamics \cite{Boyer1975,Boyer1980}, where it has been shown
that a classical charged harmonic oscillator immersed in a classical
electromagnetic zero-point field has a steady-state regime where the
phase-averaged power absorbed by the oscillator balances the phase-averaged
power radiated by the oscillator back to the zero-point field, yielding
a steady-state oscillation at the natural frequency of the oscillator
\cite{Boyer1975,Boyer1980,Puthoff1987,HuangBatelaan2013,HuangBatelaan2015,Puthoff2016}.
However, in keeping with our phenomenological approach, we will not
propose a specific mechanism for this energy exchange in ZSM, only
provisionally assume that it occurs somehow.) Accordingly, we suppose
that, in this steady-state regime, the particle undergoes undergoes
a steady-state \emph{zbw} oscillation of angular frequency $\omega_{c}$
about $\mathbf{q}_{0}$ in its IMFTRF, as characterized by the `fluctuation-dissipation'
relation, $<H>_{steady-state}=\hbar\omega_{c}=mc^{2}$, where $<H>_{steady-state}$
is the conserved random-phase-average energy associated with the steady-state
oscillation. 
\end{enumerate}
It follows then that, in the IMFTRF, the mean forward steady-state
\emph{zbw} phase change is given by 
\begin{equation}
\delta\bar{\theta}_{0+}\coloneqq\omega_{c}\delta t_{0}=\frac{mc^{2}}{\hbar}\delta t_{0},
\end{equation}
and the cumulative forward steady-state \emph{zbw} phase, obtained
from the indefinite integral of (67), is 
\begin{equation}
\bar{\theta}_{0+}=\omega_{c}t_{0}+\phi=\frac{mc^{2}}{\hbar}t_{0}+\phi_{+},
\end{equation}
where $\phi_{+}$ is the initial (forward) phase constant.

The reason for starting our analysis with the IMFTRF goes back to
the fact that, before constraining the diffusion process to simultaneous
solutions of the forward and backward Fokker-Planck equations associated
to (65-66), neither the forward nor the backward stochastic differential
equations (65-66) have well-defined time reversals. So the forward
and backward stochastic differential equations describe independent,
time-asymmetric diffusion processes in opposite time directions, and
we must start by considering the steady-state \emph{zbw} phase in
each time direction separately. We chose to start with the more intuitive
forward time direction.

For the \emph{zbw} particle in the instantaneous mean backward translational
rest frame (IMBTRF), i.e., the frame defined by $D_{*}\mathbf{q}(t)=\mathbf{b}_{*}(\mathbf{q}(t),t)=0$,
its mean backward steady-state \emph{zbw} phase change is given by
\begin{equation}
\delta\bar{\theta}_{0-}\coloneqq-\omega_{c}\delta t_{0}=-\frac{mc^{2}}{\hbar}\delta t_{0},
\end{equation}
and 
\begin{equation}
\bar{\theta}_{0-}=\left(-\omega_{c}t_{0}\right)+\phi=\left(-\frac{mc^{2}}{\hbar}t_{0}\right)+\phi_{-}.
\end{equation}

Note that, in the above construction, both the diffusion coefficient
$\nu=\hbar/2m$ and the (reduced) \emph{zbw} period $T_{c}=1/\omega_{c}=\hbar/mc^{2}$
are scaled by $\hbar$. This is consistent with our hypothesis that
the ether is the common physical cause of both the frictionless diffusion
process and the steady-state \emph{zbw} oscillation. Had we not proposed
Nelson's ether as the physical cause of the steady-state \emph{zbw}
oscillation as well as the frictionless diffusion process, the occurrence
of $\hbar$ in both of these particle properties would be inexplicable
and compromising for the plausibility of our proposed modification
of NYSM.

It should be stressed here that it is not possible to talk of the
\emph{zbw} phase in a rest frame other than the IMFTRF or IMBTRF of
the \emph{zbw} particle, as we cannot transform to a frame in which
$d\mathbf{q}(t)/dt=0$, since such an expression is undefined for
the (non-differentiable) Wiener process.

Now suppose we Lorentz transform back to the lab frame. For the forward
time direction, this corresponds to a boost of (67) by $-\mathbf{b}(\mathbf{q}(t),t)$
where $\mathbf{b}(\mathbf{q}(t),t)\neq0$. Approximating the transformation
for non-relativistic velocities so that $\gamma=1/\sqrt{\left(1-\mathbf{b}^{2}/c^{2}\right)}\approx1+\mathbf{b}^{2}/2c^{2},$
the forward steady-state \emph{zbw} phase change (67) becomes 
\begin{equation}
\begin{aligned}\delta\bar{\theta}_{+}(\mathbf{q}(t),t) & \coloneqq\frac{\omega_{c}}{mc^{2}}\mathrm{E}_{t}\left[E_{+}(D\mathbf{q}(t))\delta t-mD\mathbf{q}(t)\cdot\left(D\mathbf{q}(t)\right)\delta t\right]\\
 & =\frac{\omega_{c}}{mc^{2}}\mathrm{E}_{t}\left[E_{+}\delta t-m\mathbf{b}(\mathbf{q}(t),t)\cdot\delta\mathbf{q}_{+}(t)\right],
\end{aligned}
\end{equation}
where 
\begin{equation}
E_{+}(D\mathbf{q}(t))=mc^{2}+\frac{1}{2}m\left|D\mathbf{q}(t)\right|^{2}=mc^{2}+\frac{1}{2}m\mathbf{b}^{2},
\end{equation}
neglecting the momentum term proportional to $\mathbf{b}^{3}/c^{2}$,
and where $\delta\mathbf{q}_{+}(t)$ in (71) corresponds to the physical,
translational, mean forward displacement of the \emph{zbw} particle,
defined by 
\begin{equation}
\delta\mathbf{q}_{+}(t)=\left[D\mathbf{q}(t)\right]\delta t=\mathbf{b}(\mathbf{q}(t),t)\delta t.
\end{equation}
The first line on the right hand side of (71) is the straightforward
stochastic generalization of the Lorentz-transformed classical \emph{zbw}
phase (just as Yasue's mean action functional (19) is the straightforward
stochastic generalization of the ordinary action functional in classical
mechanics \cite{Yasue1981a}) for non-relativistic velocities. Note,
however, that the conditional expectation $\mathrm{E}_{t}[...]$ in
(71) is redundant since the right hand side of (71) involves terms
depending only on the mean forward velocity $D\mathbf{q}(t)=\mathbf{b}(\mathbf{q}(t),t)$,
where $D$ already involves taking a conditional expectation (see
the definitions (13) and (14) in section 2). However, in the more
general case of a \emph{zbw} particle in an external potential $V_{ext}$,
a case we will consider in the next section, the conditional expectation
cannot be dropped since there will be an external-potential-dependent
term in $E_{+}$ that will depend directly on $\mathbf{q}(t)$ via
$V_{ext}(\mathbf{q}(t))$. The expectation will also be useful for
giving a natural connection between the integral of the time-symmetrized
analogue of (71) (which we will introduce shortly) and Yasue's mean
action functional, as we will show later in this section.

For the backward time direction, the Lorentz transformation to the
lab frame corresponds to a boost of (69) by $-\mathbf{b}_{*}(\mathbf{q}(t),t)$
where $\mathbf{b}_{*}(\mathbf{q}(t),t)\neq0$. Then the backward steady-state
\emph{zbw} phase change (69) becomes 
\begin{equation}
\begin{aligned}\delta\bar{\theta}_{-}(\mathbf{q}(t),t) & \coloneqq\frac{\omega_{c}}{mc^{2}}\mathrm{E}_{t}\left[-E_{-}(D_{*}\mathbf{q}(t))\delta t+mD_{*}\mathbf{q}(t)\cdot\left(D_{*}\mathbf{q}(t)\right)\delta t\right]\\
 & =\frac{\omega_{c}}{mc^{2}}\mathrm{E}_{t}\left[-E_{-}\delta t+m\mathbf{b}_{*}(\mathbf{q}(t),t)\cdot\delta\mathbf{q}_{+}(t)\right],
\end{aligned}
\end{equation}
where 
\begin{equation}
E_{-}(D_{*}\mathbf{q}(t))=mc^{2}+\frac{1}{2}m\left|D_{*}\mathbf{q}(t)\right|^{2}=mc^{2}+\frac{1}{2}m\mathbf{b}_{*}^{2},
\end{equation}
and where $\delta\mathbf{q}_{-}(t)$ in (74) corresponds to the physical,
translational, mean backward displacement of the \emph{zbw} particle,
defined by 
\begin{equation}
\delta\mathbf{q}_{-}(t)=\left[D_{*}\mathbf{q}(t)\right]\delta t=\mathbf{b}_{*}(\mathbf{q}(t),t)\delta t.
\end{equation}
(Notice that $\delta\mathbf{q}_{+}(t)$ and $\delta\mathbf{q}_{-}(t)$
are not equal in general since $\delta\mathbf{q}_{+}(t)-\delta\mathbf{q}_{-}(t)=(\mathbf{b}-\mathbf{b}_{*})\delta t\neq0$
in general.) Since, at this stage, the forward and backward steady-state
\emph{zbw} phase changes, (71) and (74), are independent of one another,
each must equal $2\pi n$ when integrated along a closed loop $L$
in which both time and position change. Otherwise we will contradict
our hypothesis that, up to this point, the \emph{zbw} particle has
a well-defined steady-state phase at each point along its mean space-time
trajectory in the forward or backward time direction.

In the lab frame, the forward and backward stochastic differential
equations for the \emph{zbw} particle's translational motion are as
before 
\begin{equation}
d\mathbf{q}(t)=\mathbf{b}(\mathbf{q}(t),t)dt+d\mathbf{W}(t),
\end{equation}
and 
\begin{equation}
d\mathbf{q}(t)=\mathbf{b}_{*}(\mathbf{q}(t),t)dt+d\mathbf{W}_{*}(t),
\end{equation}
with corresponding Fokker-Planck equations 
\begin{equation}
\frac{\partial\rho(\mathbf{q},t)}{\partial t}=-\nabla\cdot\left[\mathbf{b(}\mathbf{q},t)\rho(\mathbf{q},t)\right]+\frac{\hbar}{2m}\nabla^{2}\rho(\mathbf{q},t),
\end{equation}
and 
\begin{equation}
\frac{\partial\rho(\mathbf{q},t)}{\partial t}=-\nabla\cdot\left[\mathbf{b}_{*}(\mathbf{q},t)\rho(\mathbf{q},t)\right]-\frac{\hbar}{2m}\nabla^{2}\rho(\mathbf{q},t).
\end{equation}
Restricting the diffusion process to simultaneous solutions of (79)
and (80) via 
\begin{equation}
\mathbf{v}\coloneqq\frac{1}{2}\left[\mathbf{b}+\mathbf{b}_{*}\right]=\frac{\nabla S(\mathbf{q},t)}{m}
\end{equation}
and 
\begin{equation}
\mathbf{u}\coloneqq\frac{1}{2}\left[\mathbf{b}-\mathbf{b}_{*}\right]=\frac{\hbar}{2m}\frac{\nabla\rho(\mathbf{q},t)}{\rho(\mathbf{q},t)}
\end{equation}
reduces the forward and backward Fokker-Planck equations to 
\begin{equation}
\frac{\partial\rho({\normalcolor \mathbf{q}},t)}{\partial t}=-\nabla\cdot\left[\mathbf{\frac{\nabla\mathrm{\mathit{S\mathrm{(\mathbf{q},\mathit{t})}}}}{\mathit{m}}}\rho(\mathbf{q},t)\right],
\end{equation}
with $\mathbf{b}=\mathbf{v}+\mathbf{u}$ and $\mathbf{b}_{*}=\mathbf{v}-\mathbf{u}$.
We also follow Nelson in postulating the presence of an external osmotic
potential $U(\mathbf{q},t)$ which couples to the \emph{zbw} particle
as $R(\mathbf{q},t)=\mu U(\mathbf{q},t)$, and by the same reasoning
discussed in section 2, imparts an osmotic velocity $\nabla R/m=\left(\hbar/2m\right)\nabla\rho/\rho$.
We then have $\rho=e^{2R/\hbar}$ for all times.

To obtain the 2nd-order time-symmetric mean dynamics for the translational
motion of the \emph{zbw} particle, we will use the variational principle
of Yasue. To do this, we must first define the time-symmetrized steady-state
phase change of the \emph{zbw} particle in the lab frame, via a symmetric
combination of the forward and backward steady-state \emph{zbw} phase
changes (71) and (74). This is natural to do since (71) and (74) correspond
to the same frame (the lab frame), and since (71) and (74) are no
longer independent of one another as a result of the constraints (81-82).
Taking the difference between (74) and (71), we obtain (replacing
$\delta t\rightarrow dt$, hence $\delta\mathbf{q}_{+,-}(t)\rightarrow d\mathbf{q}_{+,-}(t)$)
\begin{equation}
\begin{aligned}d\bar{\theta}(\mathbf{q}(t),t) & \coloneqq\frac{1}{2}\left[d\bar{\theta}_{+}(\mathbf{q}(t),t)-d\bar{\theta}_{-}(\mathbf{q}(t),t)\right]\\
 & =\frac{\omega_{c}}{mc^{2}}\mathrm{E}_{t}\left[E(D\mathbf{q}(t),D_{*}\mathbf{q}(t))dt-\frac{m}{2}\left(\mathbf{b}(\mathbf{q}(t),t)\cdot d\mathbf{q}_{+}(t)+\mathbf{b}_{*}(\mathbf{q}(t),t)\cdot d\mathbf{q}_{-}(t)\right)\right]\\
 & =\frac{\omega_{c}}{mc^{2}}\mathrm{E}_{t}\left[Edt-\frac{m}{2}\left(\mathbf{b}\cdot\frac{d\mathbf{q}_{+}(t)}{dt}+\mathbf{b}_{*}\cdot\frac{d\mathbf{q}_{-}(t)}{dt}\right)dt\right]\\
 & =\frac{\omega_{c}}{mc^{2}}\mathrm{E}_{t}\left[\left(E-\frac{m}{2}\left(\mathbf{b}\cdot\frac{d\mathbf{q}_{+}(t)}{dt}+\mathbf{b}_{*}\cdot\frac{d\mathbf{q}_{-}(t)}{dt}\right)\right)dt\right]\\
 & =\frac{\omega_{c}}{mc^{2}}\mathrm{E}_{t}\left[\left(E-\frac{m}{2}\left(\mathbf{b}^{2}+\mathbf{b}_{*}^{2}\right)\right)dt\right]\\
 & =\frac{\omega_{c}}{mc^{2}}\mathrm{E}_{t}\left[\left(E-\left(m\mathbf{v}\cdot\mathbf{v}+m\mathbf{u}\cdot\mathbf{u}\right)\right)dt\right]\\
 & =\frac{\omega_{c}}{mc^{2}}\mathrm{E}_{t}\left[\left(mc^{2}-\frac{1}{2}m\mathbf{v}^{2}-\frac{1}{2}m\mathbf{u}^{2}\right)dt\right],
\end{aligned}
\end{equation}
where, from (72) and (75), we have defined 
\begin{equation}
E=\frac{1}{2}\left(E_{+}+E_{-}\right)=mc^{2}+\frac{1}{2}\left[\frac{1}{2}m\mathbf{b}^{2}+\frac{1}{2}m\mathbf{b}_{*}^{2}\right]=mc^{2}+\frac{1}{2}m\mathbf{v}^{2}+\frac{1}{2}m\mathbf{u}^{2},
\end{equation}
and where we have used (73) and (76) in (84).

It is important to note that because $\bar{\theta}_{+}$ and $\bar{\theta}_{-}$
are no longer independent of one another, it is no longer the case
that $\delta\bar{\theta}_{+}$ and $\delta\bar{\theta}_{-}$ will
each equal $2\pi n$ when integrated along a closed loop $L$ in which
both time and position change. However, the consistency of our theory
does require that $\oint_{L}\delta\bar{\theta}=2\pi n$, otherwise
we would contradict our hypothesis that the \emph{zbw} particle, after
restricting to simultaneous solutions of (79) an (80), has a well-defined
and unique steady-state phase at each 3-space location it can occupy
at each time, regardless of time-direction. Note also that, without
the constraints (81-82), we would always have $\oint_{L}\delta\bar{\theta}_{+}=2\pi n$
and $\oint_{L}\delta\bar{\theta}_{-}=2\pi n$, hence $\oint_{L}\delta\bar{\theta}=0$.
In other words, a time-symmetrized ``phase'' defined from the subtractive
combination of $\bar{\theta}_{+}$ and $\bar{\theta}_{-}$, without
the constraints (81-82), would be globally single-valued instead of
single-valued up to an additive integer multiple of $2\pi$.

Now, from the last line of (84), we can integrate and define the time-symmetric
steady-state phase-principal function as 
\begin{equation}
I(\mathbf{q}(t),t)=-\hbar\bar{\theta}(\mathbf{q}(t),t)\coloneqq\mathrm{E}\left[\int_{t_{i}}^{t}\left(\frac{1}{2}m\mathbf{v}^{2}+\frac{1}{2}m\mathbf{u}^{2}-mc^{2}\right)dt'-\hbar\phi\left|\mathbf{q}(t)\right.\right],
\end{equation}
where the expectation on the right hand side is now conditional on
the Nelsonian path $\mathbf{q}(t)$. (Note that the interchangeability
of the expectation and the time integral follows from Fubini's theorem
in stochastic calculus, since the integral of the conditional expectation
and the conditional expectation of the integral are both required
to be finite quantities here \cite{Klebaner2005}.) We note that (86)
is formally identical to the \emph{W} function introduced by Yasue
in \cite{Yasue1981a}, and from which Yasue shows that the variation
$\delta W/\delta\mathbf{q}(t)$ implies the current velocity relation
(81) with \emph{W} in place of \emph{S}. The latter result also applies
to (86), given the formal identicality between \emph{I} and \emph{W},
however we will use a different approach to connect $\nabla I$ with
the current velocity (81). Also, whereas Yasue's \emph{W} function
isn't constrained to satisfy $\oint_{L}\delta W=nh$, (86) does satisfy
$\oint_{L}\delta I=nh$ since it is explicitly defined in terms of
the phase function $\bar{\theta}$.

By a slight modification of (86), we can also define the steady-state
phase-action functional 
\begin{equation}
\begin{aligned}J & \coloneqq I_{if}=\mathrm{E}\left[\int_{t_{i}}^{t_{f}}\left[\frac{1}{2}m\mathbf{v}^{2}+\frac{1}{2}m\mathbf{u}^{2}-mc^{2}\right]dt-\hbar\phi\right],\end{aligned}
\end{equation}
where $\phi$ is the initial phase constant, and where (87) differs
from (86) by the end-point at $t_{f}$ being fixed and $\mathrm{E}[...]$
being the absolute expectation. It is easily seen that (87) is just
Yasue's time-symmetric ensemble-averaged action functional, Eq. (19)
in section 2, with $V=0$, inclusion of the rest-energy term $-mc^{2}$,
and inclusion of the initial phase constant $\phi$.

Note, also, that from the second to last line of (84), we can obtain
the cumulative, time-symmetric, steady-state phase at a time \emph{t}
as 
\begin{equation}
\begin{aligned}\bar{\theta}(\mathbf{q}(t),t) & =\frac{\omega_{c}}{mc^{2}}\mathrm{E}\left[\int_{t_{i}}^{t}\left(E-\left(m\mathbf{v}\cdot\mathbf{v}+m\mathbf{u}\cdot\mathbf{u}\right)\right)dt'\left|\mathbf{q}(t)\right.\right]+\phi\\
 & =\frac{\omega_{c}}{mc^{2}}\mathrm{E}\left[\int_{t_{i}}^{t}\left(\left(E-m\mathbf{u}\cdot\mathbf{u}\right)-m\mathbf{v}\cdot\mathbf{v}\right)dt'\left|\mathbf{q}(t)\right.\right]+\phi\\
 & =\frac{\omega_{c}}{mc^{2}}\mathrm{E}\left[\int_{t_{i}}^{t}\left(H-m\mathbf{v}\cdot\mathbf{v}\right)dt'\left|\mathbf{q}(t)\right.\right]+\phi\\
 & =\frac{\omega_{c}}{mc^{2}}\mathrm{E}\left[\int_{t_{i}}^{t}\left(H-\frac{m}{4}\left(D\mathbf{q}(t')+D_{*}\mathbf{q}(t')\right)\cdot\left(D+D_{*}\right)\mathbf{q}(t')\right)dt'\left|\mathbf{q}(t)\right.\right]+\phi\\
 & =\frac{\omega_{c}}{mc^{2}}\mathrm{E}\left[\int_{t_{i}}^{t}Hdt'-\int_{\mathbf{q}(t_{i})}^{\mathbf{q}(t)}\frac{m}{2}\left(D\mathbf{q}(t')+D_{*}\mathbf{q}(t')\right)\cdot\mathrm{D}\mathbf{q}(t')\left|\mathbf{q}(t)\right.\right]+\phi,
\end{aligned}
\end{equation}
where 
\begin{equation}
H\coloneqq E-m\mathbf{u}\cdot\mathbf{u}=mc^{2}+\frac{1}{2}m\mathbf{v}^{2}-\frac{1}{2}m\mathbf{u}^{2},
\end{equation}
and where we have used the fact that $0.5\left(D+D_{*}\right)\mathbf{q}(t)=\left(\partial_{t}+\mathbf{v}\cdot\nabla\right)\mathbf{q}(t)$,
and $\mathbf{v}(\mathbf{q}(t),t)=\left(\partial_{t}+\mathbf{v}\cdot\nabla\right)\mathbf{q}(t)\eqqcolon\mathrm{D}\mathbf{q}(t)/\mathrm{D}t$,
and and $\mathrm{D}\mathbf{q}(t)=\left(\mathrm{D}\mathbf{q}(t)/\mathrm{D}t\right)dt$.
Now, given an integral curve $\mathbf{Q}(t)$ of the current velocity/momentum
field, i.e., a solution of 
\begin{equation}
m\frac{d\mathbf{Q}(t)}{dt}=m\mathbf{v}(\mathbf{Q}(t),t)=\mathbf{p}(\mathbf{Q}(t),t)=\nabla S(\mathbf{q},t)|_{\mathbf{q}=\mathbf{Q}(t)},
\end{equation}
and given that $\bar{\theta}(\mathbf{q},t)=\bar{\theta}|_{\mathbf{q}(t)=\mathbf{q}}$
is a field on 3-space representing the possible phases that the actual
\emph{zbw} particle could have at a point $\mathbf{q}$ at time \emph{t}
(up to addition of a constant), we can also evaluate $\bar{\theta}(\mathbf{q},t)$
with respect to $\mathbf{Q}(t)$, which allows us to drop the conditional
expectation (since $\mathbf{Q}(t)$ is deterministic) to obtain 
\begin{equation}
\begin{aligned}\bar{\theta}(\mathbf{Q}(t),t) & =\frac{\omega_{c}}{mc^{2}}\int_{t_{i}}^{t}\left[H-m\mathbf{v}(\mathbf{Q}(t'),t')\cdot\frac{d\mathbf{Q}(t')}{dt'}\right]dt'+\phi\\
 & =\frac{\omega_{c}}{mc^{2}}\left[\int_{t_{i}}^{t}Hdt'-\int_{\mathbf{Q}(t_{i})}^{\mathbf{Q}(t)}\mathbf{p}\cdot d\mathbf{Q}(t')\right]+\phi.
\end{aligned}
\end{equation}
Here (91) corresponds to the time-symmetrized steady-state phase of
the \emph{zbw} particle in the lab frame, evaluated along the \emph{zbw}
particle's `time-symmetric mean trajectory', where the time-symmetric
mean trajectory corresponds to an integral curve of the current velocity
field, i.e., (90). That the time-symmetric mean trajectories should
correspond to integral curves of the current velocity field can be
seen from the fact that the single-time probability density $\rho(\mathbf{q},t)$,
after imposing (81-82), is a solution of the continuity equation (83),
from which it follows that the possible mean trajectories of the \emph{zbw}
particle are the flow lines of the probability current $\rho\mathbf{v}$,
i.e., the solutions of (90) for all possible initial conditions $\mathbf{Q}(0)$.

Now, taking the total differential of the left hand side of (91) gives
\begin{equation}
d\bar{\theta}=\nabla\bar{\theta}|_{\mathbf{q}=\mathbf{Q}(t)}d\mathbf{Q}(t)+\partial_{t}\bar{\theta}|_{\mathbf{q}=\mathbf{Q}(t)}dt.
\end{equation}
This allows us to identify 
\begin{equation}
\mathbf{p}(\mathbf{Q}(t),t)=-\left(\frac{mc^{2}}{\omega_{c}}\right)\nabla\bar{\theta}|_{\mathbf{q}=\mathbf{Q}(t)}=\nabla S|_{\mathbf{q}=\mathbf{Q}(t)},
\end{equation}
where we have used (92) along with (91) and (90). Thus the current
velocity of the \emph{zbw} particle can be identified with the gradient
of the \emph{zbw} particle's time-symmetrized steady-state phase with
respect to the location of the \emph{zbw} particle at time \emph{t}
in the lab frame, given the assumption that the current velocity is
integrable, i.e., given (81) and (90). Accordingly, the $S$ function
can be identified with (91). In addition, (92) along with (91) relates
the \emph{H} function to $\bar{\theta}$ (hence \emph{S}) by 
\begin{equation}
H(\mathbf{Q}(t))=\left(\frac{mc^{2}}{\omega_{c}}\right)\partial_{t}\bar{\theta}|_{\mathbf{q}=\mathbf{Q}(t)}=-\partial_{t}S|_{\mathbf{q}=\mathbf{Q}(t)}.
\end{equation}
From (94), (93), and (91), it follows that 
\begin{equation}
\begin{aligned}S(\mathbf{Q}(t),t) & =\int_{\mathbf{Q}(t_{i})}^{\mathbf{Q}(t)}\mathbf{p}\cdot d\mathbf{Q}(t')-\int_{t_{i}}^{t}Hdt'-\hbar\phi\\
 & =\int_{t_{i}}^{t}\left[\frac{1}{2}m\mathbf{v}(\mathbf{Q}(t'),t')^{2}+\frac{1}{2}m\mathbf{u}(\mathbf{Q}(t'),t')^{2}-mc^{2}\right]dt'-\hbar\phi=I(\mathbf{Q}(t),t),
\end{aligned}
\end{equation}
and 
\begin{equation}
\oint_{L}\delta S(\mathbf{Q}(t),t)=\left(-\frac{mc^{2}}{\omega_{c}}\right)\oint_{L}\delta\bar{\theta}(\mathbf{q}(t),t)=\oint_{L}\left[\mathbf{p}\cdot\delta\mathbf{Q}(t)-H\delta t\right]=nh.
\end{equation}
We will use these last two expressions for later comparisons.

As an aside, let us recall that after restricting the forward and
backward diffusions to simultaneous solutions of (79-80), we had $\mathbf{b}=\mathbf{v}+\mathbf{u}$
and $\mathbf{b}_{*}=\mathbf{v}-\mathbf{u}$. So the IMFTRF and the
IMBTRF will not coincide since for $\mathbf{b}=\mathbf{v}+\mathbf{u}=0$
it will not generally be the case that $\mathbf{b}_{*}=\mathbf{v}-\mathbf{u}=0$.
Nevertheless, we can define an instantaneous mean (time-)symmetric
rest frame (IMSTRF) as the frame in which $\mathbf{b}+\mathbf{b}_{*}=2\mathbf{v}=0$.
In the IMSTRF, (88) or (91) or (95) reduces to $\bar{\theta}=(\omega_{c}/mc^{2})\left[\left(mc^{2}-\frac{1}{2}m\mathbf{u}^{2}\right)t+\phi\right]$,
since $\mathbf{v}=0$ and $\partial_{t}\rho=0$. This shows that the
kinetic energy term due to the osmotic velocity contributes a tiny
shift to the steady-state \emph{zbw} phase (88) or (91) or (95) in
the IMSTRF (since, in the non-relativistic regime, $\mathbf{u}^{2}/c^{2}\ll1$).

Returning now to (87), the imposition of the conservative-diffusions
constraint implies extremality of (87), which further implies (see
Appendix A) Nelson's mean acceleration equation, 
\begin{equation}
m\mathbf{a}(\mathbf{q}(t),t)=\frac{m}{2}\left[D_{*}D+DD_{*}\right]\mathbf{q}(t)=0.
\end{equation}
Computing the derivatives in (97), and using that $\mathbf{b}=\mathbf{v}+\mathbf{u}$
and $\mathbf{b}_{*}=\mathbf{v}-\mathbf{u}$, we obtain 
\begin{equation}
\begin{aligned}m\mathbf{a}(\mathbf{q}(t),t) & =m\left[\frac{\partial\mathbf{v}(\mathbf{q},t)}{\partial t}+\mathbf{v}(\mathbf{q},t)\cdot\nabla\mathbf{v}(\mathbf{q},t)-\mathbf{u}(\mathbf{q},t)\cdot\nabla\mathbf{u}(\mathbf{q},t)-\frac{\hbar}{2m}\nabla^{2}\mathbf{u}(\mathbf{q},t)\right]|_{\mathbf{q}=\mathbf{q}(t)}\\
 & =\nabla\left[\frac{\partial S(\mathbf{q},t)}{\partial t}+\frac{\left(\nabla S(\mathbf{q},t)\right)^{2}}{2m}-\frac{\hbar^{2}}{2m}\frac{\nabla^{2}\sqrt{\rho(\mathbf{q},t)}}{\sqrt{\rho(\mathbf{q},t)}}\right]|_{\mathbf{q}=\mathbf{q}(t)}=0.
\end{aligned}
\end{equation}
Integrating both sides of (98) gives the total translational energy
of the \emph{zbw} particle along the stochastic trajectory $\mathbf{q}(t)$:
\begin{equation}
\tilde{E}(\mathbf{q}(t),t)=-\frac{\partial S(\mathbf{q},t)}{\partial t}|_{\mathbf{q}=\mathbf{q}(t)}=mc^{2}+\frac{\left(\nabla S(\mathbf{q},t)\right)^{2}}{2m}|_{\mathbf{q}=\mathbf{q}(t)}-\frac{\hbar^{2}}{2m}\frac{\nabla^{2}\sqrt{\rho(\mathbf{q},t)}}{\sqrt{\rho(\mathbf{q},t)}}|_{\mathbf{q}=\mathbf{q}(t)},
\end{equation}
where we have set the integration constant equal to the \emph{zbw}
particle's rest energy. Alternatively, we can again consider integral
curves of the current velocity/momentum field, but where now the integral
curves are obtained from solutions of 
\begin{equation}
m\frac{d^{2}\mathbf{Q}(t)}{dt^{2}}=m\left(\partial_{t}\mathbf{v}+\mathbf{v}\cdot\nabla\mathbf{v}\right)|_{\mathbf{q}=\mathbf{Q}(t)}=-\nabla\left(-\frac{\hbar^{2}}{2m}\frac{\nabla^{2}\sqrt{\rho(\mathbf{q},t)}}{\sqrt{\rho(\mathbf{q},t)}}\right)|_{\mathbf{q}=\mathbf{Q}(t)},
\end{equation}
i.e., the mean acceleration equation (98), rewritten so that only
the \textbf{v}-dependent terms are kept on the left hand side. Then
we can replace $\mathbf{q}(t)$ in (99) with $\mathbf{Q}(t)$ to obtain
the total translational energy associated with the \emph{zbw} particle's
time-symmetric mean trajectory, i.e., $\tilde{E}(\mathbf{Q}(t),t)$.
Moreover, we can express the solution of (99) in terms of $\mathbf{Q}(t)$,
thereby obtaining 
\begin{equation}
\begin{aligned}S(\mathbf{Q}(t),t) & =\int_{\mathbf{Q}(t_{i})}^{\mathbf{Q}(t)}\mathbf{p}\cdot d\mathbf{Q}(t')-\int_{t_{i}}^{t}\tilde{E}dt'-\hbar\phi\\
 & =\int_{t_{i}}^{t}\left[\frac{1}{2}m\mathbf{v}(\mathbf{Q}(t'),t')^{2}-\left(-\frac{\hbar^{2}}{2m}\frac{\nabla^{2}\sqrt{\rho(\mathbf{Q}(t'),t')}}{\sqrt{\rho(\mathbf{Q}(t'),t')}}\right)-mc^{2}\right]dt'-\hbar\phi\\
 & =\int_{t_{i}}^{t}\left[\frac{1}{2}m\mathbf{v}^{2}+\frac{1}{2}m\mathbf{u}^{2}+\frac{\hbar}{2}\nabla\cdot\mathbf{u}-mc^{2}\right]dt'-\hbar\phi.
\end{aligned}
\end{equation}
We identify (101) as the conservative-diffusion-constrained, time-symmetrized,
steady-state phase (action) of the \emph{zbw} particle in the lab
frame, evaluated along an integral curve $\mathbf{Q}(t)$ obtained
from (100).

Notice that the last line of (101) differs from the last line of (95)
only by addition of the term involving $\nabla\cdot\mathbf{u}$. (The
equality between the last two lines of (101) follows from the well-known
fact that the quantum kinetic can be decomposed as $\left(-\hbar^{2}/2m\right)\rho^{-1/2}\nabla^{2}\rho^{1/2}=0.5m\mathbf{u}^{2}-\left(\hbar^{2}/4m\right)\rho^{-1}\nabla^{2}\rho$
\cite{Bohm1952I}, and by the product rule, $0.5m\mathbf{u}^{2}-\left(\hbar^{2}/4m\right)\rho^{-1}\nabla^{2}\rho=-0.5m\mathbf{u}^{2}-m\left(\hbar/2m\right)\nabla\cdot\mathbf{u}$.)

Notice also that the equation of motion for (101) differs from the
equation of motion for the classical \emph{zbw} particle phase by
the presence of the quantum kinetic entering into (98-99). The two
phases might appear to be connected by the `classical limit' $(\hbar/2m)\rightarrow0$,
but this is only a formal connection since such a limit corresponds
to deleting the presence of the ether, thereby also deleting the physical
mechanism that we hypothesize to cause the \emph{zbw} particle to
oscillate at its Compton frequency. The physically realistic `classical
limit' for (98-99) corresponds to situations where the quantum kinetic
and quantum force are negligible. Such situations will arise (as in
the dBB theory) whenever the center of mass of a system of particles
is sufficiently large and environmental decoherence is appreciable
\cite{Allori2001,Bowm2005,Oriols2016,Derakhshani2017b}.

Inasmuch as (101) is a well-defined phase function of the \emph{zbw}
particle's time-symmetric mean trajectory $\mathbf{Q}(t)$ in the
lab frame (because it was derived from applying the variational principle
to (87), the latter of which was defined in terms of (84), which we
argued must satisfy $\oint_{L}\delta\bar{\theta}=2\pi n$), if we
integrate $\delta S(\mathbf{Q}(t),t)$ around a closed loop $L$ in
which time and position may change, we will have 
\begin{equation}
\oint_{L}\delta S(\mathbf{Q}(t),t)=\oint_{L}\left[\mathbf{p}\cdot\delta\mathbf{Q}(t)-\tilde{E}\delta t\right]=nh,
\end{equation}
and for a special loop in which time is held fixed, 
\begin{equation}
\oint_{L}\nabla S|_{\mathbf{q}=\mathbf{Q}(t)}\cdot\delta\mathbf{Q}(t)=\oint_{L}\mathbf{p}\cdot\delta\mathbf{Q}(t)=nh.
\end{equation}
Otherwise, we would contradict our hypothesis that the \emph{zbw}
particle still has a well-defined, time-symmetrized, steady-state
phase at each 3-space location it can occupy along a mean trajectory
$\mathbf{Q}(t)$ in either time direction, after the constraint of
conservative diffusions has been imposed. (Notice that (102) differs
from (96) by $\tilde{E}$ replacing $H$, and that $\tilde{E}-H=-(\hbar/2)\nabla\cdot\mathbf{u}$.)
If we also consider the time-symmetrized steady-state phase field,
$S(\mathbf{q},t)$, which is a field over the possible locations of
the actual \emph{zbw} particle (as described in section 4.2), then
by applying the same physical reasoning above to each possible initial
position that the \emph{zbw} particle can occupy, it follows that
the net change of the phase field along any mathematical loop in space
(with time held fixed) will be 
\begin{equation}
\oint_{L}dS(\mathbf{q},t)=\oint_{L}\mathbf{p}\cdot d\mathbf{q}=nh.
\end{equation}
(The justification for (104) where $\rho=0$ is discussed in section
5.2, since such ``nodal points'' commonly arise in the presence
of bound states.)

The total energy field $\tilde{E}(\mathbf{q},t)$ will correspondingly
be given by (99) when $\mathbf{Q}(t)$ is replaced by $\mathbf{q}$.
So with (104), (99), and (83), we can construct the 1-particle Schrödinger
equation, 
\begin{equation}
i\hbar\frac{\partial\psi(\mathbf{q},t)}{\partial t}=-\frac{\hbar^{2}}{2m}\nabla^{2}\psi(\mathbf{q},t)+mc^{2}\psi(\mathbf{q},t),
\end{equation}
where $\psi(\mathbf{q},t)=\sqrt{\rho(\mathbf{q},t)}e^{iS(\mathbf{q},t)/\hbar}$
is a single-valued wave function as a result of (104). As in the classical
case, the constant $C=\hbar\phi$ will contribute a global phase factor
$e^{iC/\hbar}$ which cancels out from both sides of (105). We thereby
have a formulation of free-particle ZSM that recovers the usual free-particle
Schrödinger equation.

\subsection{One particle interacting with external fields}

Suppose again that the particle undergoes a steady-state \emph{zbw}
oscillation in the IMFTRF, but now carries charge $e$ so that it
is a classical charged harmonic oscillator of some type (subject again
to the hypothetical constraint of no electromagnetic radiation emitted
when there is no translational motion; or the constraint that the
oscillation of the charge is radially symmetric so that there is no
net energy radiated; or, if the ether turns out to be electromagnetic
in nature as Nelson suggested \cite{Nelson1985}, then that the steady-state
\emph{zbw} oscillation is due to a balancing between the random-phase-averaged
electromagnetic energy absorbed from the charged harmonic oscillator's
driven oscillation, and the random-phase-averaged electromagnetic
energy radiated back to the ether, much like in stochastic electrodynamics
\cite{Boyer1975,Boyer1980,Puthoff1987,HuangBatelaan2013,HuangBatelaan2015,Puthoff2016}).
Then, in the presence of an external electric potential $\Phi_{e}(\mathbf{q}_{0}(t_{0}),t_{0})=\mathbf{E}_{ext}(\mathbf{q}_{0}(t_{0}),t_{0})\cdot\mathbf{q}_{0}(t_{0})$,
where \textbf{$\mathbf{q}_{0}(t_{0})$} is the positional displacement
of the \emph{zbw} particle in some arbitrary direction from the field
source (again making the point-like approximation for $|\mathbf{q}_{0}|\gg\lambda_{c}$)
and satisfies the forward stochastic differential equation (77) with
$\mathbf{b}=0$, the\emph{ zbw} phase change in this IMFTRF is shifted
by 
\begin{equation}
\delta\bar{\theta}_{0+}=\mathrm{E}_{t}\left[\left(\omega_{c}+\varepsilon(\mathbf{q}_{0}(t_{0}),t_{0})\right)\delta t_{0}\right]=\frac{1}{\hbar}\left(mc^{2}\delta t_{0}+\mathrm{E}_{t}\left[e\Phi_{e}(\mathbf{q}_{0}(t_{0}),t_{0})\delta t_{0}\right]\right),
\end{equation}
where $\varepsilon(\mathbf{q}_{0}(t_{0}),t_{0})=\omega_{c}\left(e/mc^{2}\right)\Phi_{e}(\mathbf{q}_{0}(t_{0}),t_{0})$.
Direct integration gives 
\begin{equation}
\begin{aligned}\bar{\theta}_{0+} & =\mathrm{E}\left[\int_{t_{a}}^{t_{0}}\left(\omega_{c}+\varepsilon(\mathbf{q}_{0}(t'_{0}),t'_{0})\right)dt'_{0}\left|\mathbf{q}_{0}(t_{0})\right.\right]\\
 & =\frac{1}{\hbar}\left(mc^{2}t_{0}+\mathrm{E}\left[e\int_{t_{a}}^{t_{0}}\Phi_{e}(\mathbf{q}_{0}(t'_{0}),t'_{0})dt'_{0}\left|\mathbf{q}_{0}(t_{0})\right.\right]\right)+\phi.
\end{aligned}
\end{equation}
In the IMBTRF, 
\begin{equation}
\delta\bar{\theta}_{0-}=-\mathrm{E}_{t}\left[\left(\omega_{c}+\varepsilon(\mathbf{q}_{0}(t_{0}),t_{0})\right)\delta t_{0}\right]=-\frac{1}{\hbar}\left(mc^{2}\delta t_{0}+\mathrm{E}_{t}\left[e\Phi_{e}(\mathbf{q}_{0}(t_{0}),t_{0})\delta t_{0}\right]\right).
\end{equation}
Direct integration gives 
\begin{equation}
\begin{aligned}\bar{\theta}_{0-} & =-\mathrm{E}\left[\int_{t_{a}}^{t_{0}}\left(\omega_{c}+\varepsilon(\mathbf{q}_{0}(t'_{0}),t'_{0})\right)dt'_{0}\left|\mathbf{q}_{0}(t_{0})\right.\right]\\
 & =-\frac{1}{\hbar}\left(mc^{2}t_{0}+\mathrm{E}\left[e\int_{t_{a}}^{t_{0}}\Phi_{e}(\mathbf{q}_{0}(t'_{0}),t'_{0})dt'_{0}\left|\mathbf{q}_{0}(t_{0})\right.\right]\right)+\phi.
\end{aligned}
\end{equation}
Now suppose we Lorentz transform back to the lab frame. For the forward
time direction, this corresponds to a boost of (106) by $-\mathbf{b}(\mathbf{q}(t),t)$
where $\mathbf{b}(\mathbf{q}(t),t)\neq0$. Approximating the transformation
for non-relativistic velocities so that $\gamma=1/\sqrt{\left(1-\mathbf{b}^{2}/c^{2}\right)}\approx1+\mathbf{b}^{2}/2c^{2},$
(106) becomes 
\begin{equation}
\delta\bar{\theta}_{+}(\mathbf{q}(t),t)=\frac{\omega_{c}}{mc^{2}}\mathrm{E}_{t}\left[E_{+}(\mathbf{q}(t),D\mathbf{q}(t),t)\delta t-m\mathbf{b}(\mathbf{q}(t),t)\cdot\delta\mathbf{q}_{+}(t)\right],
\end{equation}
where 
\begin{equation}
E_{+}(\mathbf{q}(t),D\mathbf{q}(t),t)=mc^{2}+\frac{1}{2}m\mathbf{b}^{2}+e\Phi_{e},
\end{equation}
neglecting the momentum term proportional to $\mathbf{b}^{3}/c^{2}$.
Again we take $\delta\mathbf{q}_{+}(t)$ to correspond to (73). For
the backward time direction, we have a boost of (108) by $-\mathbf{b}_{*}(\mathbf{q}(t),t)$
where $\mathbf{b}_{*}(\mathbf{q}(t),t)\neq0$, hence 
\begin{equation}
\delta\bar{\theta}_{-}(\mathbf{q}(t),t)=\frac{\omega_{c}}{mc^{2}}\mathrm{E}_{t}\left[-E_{-}(\mathbf{q}(t),D_{*}\mathbf{q}(t),t)\delta t+m\mathbf{b}_{*}(\mathbf{q}(t),t)\cdot\delta\mathbf{q}_{-}(t)\right],
\end{equation}
where 
\begin{equation}
E_{-}(\mathbf{q}(t),D_{*}\mathbf{q}(t),t)=mc^{2}+\frac{1}{2}m\mathbf{b}_{*}^{2}+e\Phi_{e}.
\end{equation}
Again we take $\delta\mathbf{q}_{-}(t)$ to correspond to (76).

As in the free particle case, at this stage, the forward and backward
steady-state \emph{zbw} phase changes, (110) and (112), are independent
of one another. So both (110) and (112) must equal $2\pi n$ when
integrated along a closed loop $L$ in which both time and position
change. Otherwise we will contradict our hypothesis that, up to this
point, the \emph{zbw} particle has a well-defined mean forward or
backward steady-state phase at each point along its mean forward or
backward space-time trajectory.

In the lab frame, the forward and backward stochastic differential
equations for the translational motion are once again 
\begin{equation}
d\mathbf{q}(t)=\mathbf{b}(\mathbf{q}(t),t)+d\mathbf{W}(t),
\end{equation}
and 
\begin{equation}
d\mathbf{q}(t)=\mathbf{b}_{*}(\mathbf{q}(t),t)+d\mathbf{W}_{*}(t),
\end{equation}
with corresponding Fokker-Planck equations 
\begin{equation}
\frac{\partial\rho(\mathbf{q},t)}{\partial t}=-\nabla\cdot\left[\mathbf{b(}\mathbf{q},t)\rho(\mathbf{q},t)\right]+\frac{\hbar}{2m}\nabla^{2}\rho(\mathbf{q},t),
\end{equation}
and 
\begin{equation}
\frac{\partial\rho(\mathbf{q},t)}{\partial t}=-\nabla\cdot\left[\mathbf{b}_{*}(\mathbf{q},t)\rho(\mathbf{q},t)\right]-\frac{\hbar}{2m}\nabla^{2}\rho(\mathbf{q},t).
\end{equation}
Let us now suppose that an external magnetic field $\mathbf{B}_{ext}(\mathbf{q},t)=\nabla\times\mathbf{A}_{ext}(\mathbf{q},t)$
is also present. Then, restricting ourselves to simultaneous solutions
of (116-117) via 
\begin{equation}
\mathbf{v}\coloneqq\frac{1}{2}\left[\mathbf{b}+\mathbf{b}_{*}\right]=\frac{\nabla S}{m}-\frac{e}{mc}\mathbf{A}_{ext}
\end{equation}
and 
\begin{equation}
\mathbf{u}\coloneqq\frac{1}{2}\left[\mathbf{b}-\mathbf{b}_{*}\right]=\frac{\hbar}{2m}\frac{\nabla\rho}{\rho}
\end{equation}
entails that (116-117) reduce to 
\begin{equation}
\frac{\partial\rho}{\partial t}=-\nabla\cdot\left[\left(\mathbf{\frac{\nabla\mathrm{\mathit{S}}}{\mathit{m}}}-\frac{e}{mc}\mathbf{A}_{ext}\right)\rho\right].
\end{equation}
We can then write $\mathbf{b}'=\mathbf{v}'+\mathbf{u}$ and $\mathbf{b}'_{*}=\mathbf{v}'-\mathbf{u}$,
where we recall that $\mathbf{v}'=\mathbf{v}+(e/mc)\mathbf{A}_{ext}$,
implying $\mathbf{b}=\mathbf{b}'-(e/mc)\mathbf{A}_{ext}$ and $\mathbf{b}_{*}=\mathbf{b}'_{*}-(e/mc)\mathbf{A}_{ext}$.
Once again the osmotic potential $R(\mathbf{q},t)=\mu U(\mathbf{q},t)$
imparts to the particle an osmotic velocity $\nabla R/m=\left(\hbar/2m\right)\nabla\rho/\rho$
(see section 2), implying $\rho=e^{2R/\hbar}$ for all times.

As in the free particle case, we can obtain the 2nd-order time-symmetric
mean dynamics from Yasue's variational principle.

Since (110) and (112) correspond to the same (lab) frame and are no
longer independent because of (118-119), it is natural to define the
time-symmetric steady-state \emph{zbw} particle phase in the lab frame
by taking the difference between (110) and (112) (under the replacements
$\mathbf{b}\rightarrow\mathbf{b}'$ and $\mathbf{b}_{*}\rightarrow\mathbf{b}'_{*}$
in the mean forward and mean backward momentum contributions to the
phases): 
\begin{equation}
\begin{aligned}d\bar{\theta}(\mathbf{q}(t),t) & \coloneqq\frac{1}{2}\left[d\bar{\theta}_{+}(\mathbf{q}(t),t)-d\bar{\theta}_{-}(\mathbf{q}(t),t)\right]\\
 & =\frac{\omega_{c}}{mc^{2}}\mathrm{E}_{t}\left[E(\mathbf{q}(t),D\mathbf{q}(t),D_{*}\mathbf{q}(t),t)dt-\frac{m}{2}\left(\mathbf{b}'(\mathbf{q}(t),t)\cdot d\mathbf{q}_{+}(t)+\mathbf{b}'_{*}(\mathbf{q}(t),t)\cdot d\mathbf{q}_{-}(t)\right)\right]+\phi\\
 & =\frac{\omega_{c}}{mc^{2}}\mathrm{E}_{t}\left[Edt-\frac{m}{2}\left(\mathbf{b}'\cdot\frac{d\mathbf{q}_{+}(t)}{dt}+\mathbf{b}'_{*}\cdot\frac{d\mathbf{q}_{-}(t)}{dt}\right)dt\right]+\phi\\
 & =\frac{\omega_{c}}{mc^{2}}\mathrm{E}_{t}\left[\left(E-\frac{m}{2}\left(\mathbf{b}'\cdot\frac{d\mathbf{q}_{+}(t)}{dt}+\mathbf{b}'_{*}\cdot\frac{d\mathbf{q}_{-}(t)}{dt}\right)\right)dt\right]+\phi\\
 & =\frac{\omega_{c}}{mc^{2}}\mathrm{E}_{t}\left[\left(E-\frac{m}{2}\left(\mathbf{b}'\cdot\mathbf{b}+\mathbf{b}'_{*}\cdot\mathbf{b}_{*}\right)\right)dt\right]+\phi\\
 & =\frac{\omega_{c}}{mc^{2}}\mathrm{E}_{t}\left[\left(E-\frac{m}{2}\left(\mathbf{b}{}^{2}+\frac{e}{mc}\mathbf{b}\cdot\mathbf{A}_{ext}+\mathbf{b}_{*}^{2}+\frac{e}{mc}\mathbf{b}_{*}\cdot\mathbf{A}_{ext}\right)\right)dt\right]+\phi\\
 & =\frac{\omega_{c}}{mc^{2}}\mathrm{E}_{t}\left[\left(E-\frac{m}{2}\left(\mathbf{b}{}^{2}+\mathbf{b}_{*}^{2}\right)-\frac{e}{c}\left(\frac{\mathbf{b}+\mathbf{b}_{*}}{2}\right)\cdot\mathbf{A}_{ext}\right)dt\right]+\phi\\
 & =\frac{\omega_{c}}{mc^{2}}\mathrm{E}_{t}\left[\left(E-\left(m\mathbf{v}\cdot\mathbf{v}+m\mathbf{u}\cdot\mathbf{u}\right)-\frac{e}{c}\mathbf{v}\cdot\mathbf{A}_{ext}\right)dt\right]+\phi\\
 & =\frac{\omega_{c}}{mc^{2}}\mathrm{E}_{t}\left[\left(mc^{2}+e\Phi_{e}-\frac{1}{2}m\mathbf{v}^{2}-\frac{1}{2}m\mathbf{u}^{2}-\frac{e}{c}\mathbf{v}\cdot\mathbf{A}_{ext}\right)dt\right]+\phi.
\end{aligned}
\end{equation}
where, using (111) and (113), along with the constraints (118) and
(119), we have defined 
\begin{equation}
\begin{aligned}E(\mathbf{q}(t),D\mathbf{q}(t),D_{*}\mathbf{q}(t),t) & =mc^{2}+\frac{1}{2}\left[\frac{1}{2}m\mathbf{b}^{2}+\frac{1}{2}m\mathbf{b}_{*}^{2}\right]+e\Phi_{e}\\
 & =mc^{2}+\frac{1}{2}m\mathbf{v}^{2}+\frac{1}{2}m\mathbf{u}^{2}+e\Phi_{e}.
\end{aligned}
\end{equation}

As in the free particle case, the consistency of our theory requires
that the time-symmetrized steady-state \emph{zbw} phase change of
the \emph{zbw} particle in the lab frame, (121), satisfies $\oint_{L}\delta\bar{\theta}=2\pi n$.
Otherwise we would contradict our hypothesis that the \emph{zbw} particle,
under the time-symmetric constraints (118-119), has a well-defined
and unique steady-state phase at each 3-space location it can occupy
at each time, regardless of time direction.

Using the integral of (121) in the definition of the steady-state
phase-principal function 
\begin{equation}
I=-\frac{mc^{2}}{\omega_{c}}\bar{\theta}=\mathrm{E}\left[\int_{t_{i}}^{t}\left(\frac{1}{2}m\mathbf{v}^{2}+\frac{1}{2}m\mathbf{u}^{2}+\frac{e}{c}\mathbf{v}\cdot\mathbf{A}_{ext}-mc^{2}-e\Phi_{e}\right)dt'\left|\mathbf{q}(t)\right.\right]-\hbar\phi,
\end{equation}
we can define the steady-state phase-action functional as 
\begin{equation}
\begin{aligned}J & =I_{if}=\mathrm{E}\left[\int_{t_{i}}^{t_{f}}\left(\frac{1}{2}m\mathbf{v}^{2}+\frac{1}{2}m\mathbf{u}^{2}+\frac{e}{c}\mathbf{v}\cdot\mathbf{A}_{ext}-mc^{2}-e\Phi_{e}\right)dt'\right]-\hbar\phi.\end{aligned}
\end{equation}
Equation (124) is just Yasue's mean action functional, Eq. (145) in
Appendix A, but with the inclusion of the rest-energy term $-mc^{2}$
and the time-symmetrized initial phase constant $\phi$.

Note, also, that from the second to last line of (121), we can write
the cumulative, time-symmetric, steady-state phase at a time \emph{t}
as 
\begin{equation}
\begin{aligned}\bar{\theta}(\mathbf{q}(t),t) & =\frac{\omega_{c}}{mc^{2}}\mathrm{E}\left[\int_{t_{i}}^{t}\left(E-\left(m\mathbf{v}\cdot\mathbf{v}+m\mathbf{u}\cdot\mathbf{u}\right)-\frac{e}{c}\mathbf{v}\cdot\mathbf{A}_{ext}\right)dt'\left|\mathbf{q}(t)\right.\right]+\phi\\
 & =\frac{\omega_{c}}{mc^{2}}\mathrm{E}\left[\int_{t_{i}}^{t}\left(\left(E-m\mathbf{u}\cdot\mathbf{u}\right)-m\mathbf{v}\cdot\mathbf{v}-\frac{e}{c}\mathbf{v}\cdot\mathbf{A}_{ext}\right)dt'\left|\mathbf{q}(t)\right.\right]+\phi\\
 & =\frac{\omega_{c}}{mc^{2}}\mathrm{E}\left[\int_{t_{i}}^{t}\left(H-m\mathbf{v}\cdot\mathbf{v}-\frac{e}{c}\mathbf{v}\cdot\mathbf{A}_{ext}\right)dt'\left|\mathbf{q}(t)\right.\right]+\phi\\
 & =\frac{\omega_{c}}{mc^{2}}\mathrm{E}\left[\int_{t_{i}}^{t}Hdt'-\int_{\mathbf{q}(t_{i})}^{\mathbf{q}(t)}\left(m\mathbf{v}(\mathbf{q}(t'),t')+\frac{e}{c}\mathbf{A}_{ext}(\mathbf{q}(t'),t')\right)\cdot\mathrm{D}\mathbf{q}(t')\left|\mathbf{q}(t)\right.\right]+\phi,
\end{aligned}
\end{equation}
where 
\begin{equation}
H\coloneqq E-m\mathbf{u}\cdot\mathbf{u}=mc^{2}+\frac{1}{2}m\mathbf{v}^{2}-\frac{1}{2}m\mathbf{u}^{2}+e\Phi_{e}.
\end{equation}
Now, given an integral curve $\mathbf{Q}(t)$ obtained from 
\begin{equation}
m\frac{d\mathbf{Q}(t)}{dt}=\mathbf{p}(\mathbf{Q}(t),t)=\nabla S(\mathbf{q},t)|_{\mathbf{q}=\mathbf{Q}(t)}-\frac{e}{c}\mathbf{A}_{ext}(\mathbf{Q}(t),t),
\end{equation}
we can replace (125) with 
\begin{equation}
\begin{aligned}\bar{\theta}(\mathbf{Q}(t),t) & =\frac{\omega_{c}}{mc^{2}}\int_{t_{i}}^{t}\left(H-m\mathbf{v}\cdot\frac{d\mathbf{Q}(t')}{dt'}-\frac{e}{c}\frac{d\mathbf{Q}(t')}{dt'}\cdot\mathbf{A}_{ext}(\mathbf{Q}(t'),t')\right)dt'+\phi\\
 & =\frac{\omega_{c}}{mc^{2}}\left[\int_{t_{i}}^{t}Hdt'-\int_{\mathbf{Q}(t_{i})}^{\mathbf{Q}(t)}\left(\mathbf{p}+\frac{e}{c}\mathbf{A}_{ext}\right)\cdot d\mathbf{Q}(t')\right]+\phi.
\end{aligned}
\end{equation}
The total differential of the left hand side of (128) gives 
\begin{equation}
d\bar{\theta}=\nabla\bar{\theta}|_{\mathbf{q}=\mathbf{Q}(t)}d\mathbf{Q}(t)+\partial_{t}\bar{\theta}|_{\mathbf{q}=\mathbf{Q}(t)}dt.
\end{equation}
Hence, 
\begin{equation}
\mathbf{p}(\mathbf{Q}(t),t)+\frac{e}{c}\mathbf{A}_{ext}(\mathbf{Q}(t),t)=-\left(\frac{mc^{2}}{\omega_{c}}\right)\nabla\bar{\theta}|_{\mathbf{q}=\mathbf{Q}(t)}=\nabla S|_{\mathbf{q}=\mathbf{Q}(t)}.
\end{equation}
Thus the current velocity, plus the correction due to the external
vector potential, corresponds the gradient of the \emph{zbw} particle's
time-symmetrized steady-state phase at the location of the \emph{zbw}
particle, and $S$ can again be identified with the time-symmetrized
steady-state action/phase function of the \emph{zbw} particle in the
lab frame. Along with 
\begin{equation}
H(\mathbf{Q}(t),t)=\left(\frac{mc^{2}}{\omega_{c}}\right)\partial_{t}\bar{\theta}|_{\mathbf{q}=\mathbf{Q}(t)}=-\partial_{t}S|_{\mathbf{q}=\mathbf{Q}(t)},
\end{equation}
it follows that 
\begin{equation}
\begin{aligned}S(\mathbf{Q}(t),t) & =\int_{t_{i}}^{t}\left(\mathbf{p}+\frac{e}{c}\mathbf{A}_{ext}\right)\cdot d\mathbf{Q}(t')-\int_{t_{i}}^{t}Hdt'-\hbar\phi\\
 & =\int_{t_{i}}^{t}\left[\frac{1}{2}m\mathbf{v}^{2}+\frac{1}{2}m\mathbf{u}^{2}+\frac{e}{c}\mathbf{v}\cdot\mathbf{A}_{ext}-mc^{2}-e\Phi_{e}\right]dt'-\hbar\phi=I(\mathbf{Q}(t),t),
\end{aligned}
\end{equation}
and 
\begin{equation}
\oint_{L}\delta S(\mathbf{Q}(t),t)=\left(-\frac{mc^{2}}{\omega_{c}}\right)\oint_{L}\delta\bar{\theta}(\mathbf{q}(t),t)=\oint_{L}\left[\mathbf{p}'\cdot\delta\mathbf{Q}(t)-H\delta t\right]=nh.
\end{equation}

Recall that after restricting the forward and backward diffusions
to simultaneous solutions of (116-117), we have $\mathbf{b}=\mathbf{v}+\mathbf{u}$
and $\mathbf{b}_{*}=\mathbf{v}-\mathbf{u}$. So the IMFTRF and the
IMBTRF will not coincide since, for $\mathbf{b}=\mathbf{v}+\mathbf{u}=0$,
it will generally not be the case that $\mathbf{b}_{*}=\mathbf{v}-\mathbf{u}=0$.
This motivates defining an instantaneous mean (time-)symmetric rest
frame (IMSTRF) as the frame in which $\mathbf{b}+\mathbf{b}_{*}=2\mathbf{v}=0$.
In the IMSTRF, (128) reduces to $\bar{\theta}=(\omega_{c}/mc^{2})\left[\left(mc^{2}-\frac{1}{2}m\mathbf{u}^{2}\right)t+\int_{t_{i}}^{t}e\Phi_{e}(\mathbf{Q}_{0},t')dt'\right]+\phi$,
since $\mathbf{v}=0$ and $\partial_{t}\rho=0$. So the external potential
contributes a tiny shift to the time-symmetrized steady-state \emph{zbw}
phase in the IMSTRF, along with the kinetic energy term involving
the osmotic velocity.

Applying the conservative diffusion constraint to the steady-state
phase/action functional (124), we recover the mean acceleration equation
\begin{equation}
m\mathbf{a}(\mathbf{q}(t),t)=\frac{m}{2}\left[D_{*}D+DD_{*}\right]\mathbf{q}(t)=e\left[-\frac{1}{c}\frac{\partial\mathbf{A}_{ext}}{\partial t}-\nabla\Phi_{e}+\frac{\mathbf{v}}{c}\times\mathbf{B}_{ext}\right]|_{\mathbf{q}=\mathbf{q}(t)}.
\end{equation}
Applying the mean derivatives in (133), we find 
\begin{equation}
\begin{aligned}m\mathbf{a}(\mathbf{q}(t),t) & =m\left[\frac{\partial\mathbf{v}}{\partial t}+\mathbf{v}\cdot\nabla\mathbf{v}-\mathbf{u}\cdot\nabla\mathbf{u}-\frac{\hbar}{2m}\nabla^{2}\mathbf{u}\right]|_{\mathbf{q}=\mathbf{q}(t)}\\
 & =e\left[-\frac{1}{c}\frac{\partial\mathbf{A}_{ext}}{\partial t}-\nabla\Phi_{e}+\frac{\mathbf{v}}{c}\times\mathbf{B}_{ext}\right]|_{\mathbf{q}=\mathbf{q}(t)}.
\end{aligned}
\end{equation}
Integrating both sides gives 
\begin{equation}
\tilde{E}(\mathbf{q}(t),t)=-\frac{\partial S(\mathbf{q},t)}{\partial t}|_{\mathbf{q}=\mathbf{q}(t)}=mc^{2}+\left[\frac{\left(\nabla S-\frac{e}{c}\mathbf{A}_{ext}\right)^{2}}{2m}+e\Phi_{e}-\frac{\hbar^{2}}{2m}\frac{\nabla^{2}\sqrt{\rho}}{\sqrt{\rho}}\right]|_{\mathbf{q}=\mathbf{q}(t)},
\end{equation}
where we have fixed the integration constant equal to the particle
rest energy. Alternatively, we can again consider integral curves
of the current velocity/momentum field, but where now the integral
curves are obtained from solutions of 
\begin{equation}
\begin{aligned}m\frac{d^{2}\mathbf{Q}(t)}{dt^{2}} & =m\left(\partial_{t}\mathbf{v}+\mathbf{v}\cdot\nabla\mathbf{v}\right)|_{\mathbf{q}=\mathbf{Q}(t)}\\
 & =-\nabla\left(-\frac{\hbar^{2}}{2m}\frac{\nabla^{2}\sqrt{\rho(\mathbf{q},t)}}{\sqrt{\rho(\mathbf{q},t)}}\right)|_{\mathbf{q}=\mathbf{Q}(t)}+e\left[-\frac{1}{c}\partial_{t}\mathbf{A}_{ext}-\nabla\Phi_{e}+\frac{\mathbf{v}}{c}\times\mathbf{B}_{ext}\right]|_{\mathbf{q}=\mathbf{Q}(t)},
\end{aligned}
\end{equation}
i.e., the mean acceleration equation (98), rewritten so that only
the \textbf{v}-dependent terms are kept on the left hand side. Then
we can replace $\mathbf{q}(t)$ in (136) with $\mathbf{Q}(t)$ to
obtain $\tilde{E}(\mathbf{Q}(t),t)$. The corresponding general solution,
i.e., the time-symmetrized steady-state phase/action of the \emph{zbw}
particle in the lab frame, after having imposed the conservative diffusion
constraint on (124), is of the form 
\begin{equation}
\begin{aligned}S(\mathbf{Q}(t),t) & =\int_{\mathbf{Q}(t_{i})}^{\mathbf{Q}(t)}\left(\mathbf{p}+\frac{e}{c}\mathbf{A}_{ext}\right)\cdot d\mathbf{Q}(t')-\int_{t_{i}}^{t}\tilde{E}dt'-\hbar\phi\\
 & =\int_{t_{i}}^{t}\left[\frac{1}{2}m\mathbf{v}^{2}-\left(-\frac{\hbar^{2}}{2m}\frac{\nabla^{2}\sqrt{\rho}}{\sqrt{\rho}}\right)+\frac{e}{c}\mathbf{v}\cdot\mathbf{A}_{ext}-mc^{2}-e\Phi_{e}\right]dt'-\hbar\phi\\
 & =\int_{t_{i}}^{t}\left[\frac{1}{2}m\mathbf{v}^{2}+\frac{1}{2}m\mathbf{u}^{2}+\frac{\hbar}{2}\nabla\cdot\mathbf{u}+\frac{e}{c}\mathbf{v}\cdot\mathbf{A}_{ext}-mc^{2}-e\Phi_{e}\right]dt'-\hbar\phi.
\end{aligned}
\end{equation}
Notice that the last line of (138) differs from the last line of (132)
only by addition of the term involving $\nabla\cdot\mathbf{u}$.

As also in the free particle case, the equation of motion for (138)
differs from the equation of motion for the classical \emph{zbw} particle
phase by the presence of the quantum kinetic in (135-136). Our earlier
discussion of the quantum-classical correspondence applies here as
well.

Insofar as (138) is a well-defined phase function, if we integrate
$\delta S(\mathbf{Q}(t),t)$ around a closed loop $L$ in which time
and position may change, we will have 
\begin{equation}
\oint_{L}\delta S(\mathbf{Q}(t),t)=\oint_{L}\left[\mathbf{p}'\cdot\delta\mathbf{Q}(t)-\tilde{E}\delta t\right]=nh,
\end{equation}
and for a special loop in which time is held fixed, 
\begin{equation}
\oint_{L}\delta S(\mathbf{Q}(t))=\oint_{L}\nabla S|_{\mathbf{q}=\mathbf{Q}(t)}\cdot\delta\mathbf{Q}(t)=\oint_{L}\mathbf{p}'\cdot\delta\mathbf{Q}(t)=nh.
\end{equation}
Considering also the \emph{zbw} phase field $S(\mathbf{q},t)$, which
we recall is a field over the possible locations of the actual \emph{zbw}
particle, and applying the same physical reasoning above to each possible
initial position that the \emph{zbw} particle can occupy, it follows
that the net phase change along any mathematical loop in space (with
time held fixed) will be given by 
\begin{equation}
\oint_{L}\nabla S\cdot d\mathbf{q}=\oint_{L}\mathbf{p}'\cdot d\mathbf{q}=nh.
\end{equation}
The corresponding total energy field $E(\mathbf{q},t)$ is given by
(136) when $\mathbf{Q}(t)$ is replaced by $\mathbf{q}$. From (141),
(136), and (120), we can then construct the 1-particle Schrödinger
equation in external fields as 
\begin{equation}
i\hbar\frac{\partial\psi}{\partial t}=\frac{\left[-i\hbar\nabla-\frac{e}{c}\mathbf{A}_{ext}\right]^{2}}{2m}\psi+e\Phi_{e}\psi+mc^{2}\psi,
\end{equation}
where $\psi(\mathbf{q},t)=\sqrt{\rho(\mathbf{q},t)}e^{iS(\mathbf{q},t)/\hbar}$
is a single-valued wave function as a consequence of (141).

At this point, it is worth observing an important difference between
the (time-symmetrized steady-state \emph{zbw}) phase field evolving
by (136) and the classical \emph{zbw} phase field evolving by Eq.
(61) in section 4.4. In the former case, the nonlinear coupling to
the density $\rho$ via the quantum kinetic implies that, at nodal
points (i.e., where $\rho=\psi=0$), such as found in excited states
of the hydrogen atom or quantum harmonic oscillator, the phase field
develops a singularity where both $\mathbf{v}=\nabla S$ and $\mathbf{u}=\left(\hbar/2m\right)\nabla\ln\rho$
diverge. Moreover, (141) implies that the phase field along a closed
loop \emph{L} undergoes a discontinuous jump of magnitude $nh$ if
the loop happens to cross a nodal point. Neither of these observations
are inconsistent with our hypothesis that the steady-state phase of
the actual \emph{zbw} particle is a well-defined function of the actual
particle's mean space-time trajectory (or any mean space-time trajectory
it can potentially realize), since it can be readily shown that the
particle's actual (mean or stochastic) trajectory never hits a nodal
point \cite{Nelson1966,Nelson1985,Blanchard1987,Wallstrom1990,Holland1993}.
\footnote{A simple proof \cite{Holland1993} of this for the actual mean trajectory
can be given as follows. First, note that the actual particle's initial
mean position, $\mathbf{q}(0)$, can never be at nodal points (since
this would contradict the physical meaning of $\rho$ as the probability
density for the particle to be at position $\mathbf{q}$ at time $t$).
Now, rewrite $\partial_{t}\rho=-\nabla\cdot\left(\mathbf{v}\rho\right)$
as $\left(\partial_{t}+\mathbf{v}\cdot\nabla\right)\rho=-\rho\nabla\cdot\mathbf{v}$.
Along the actual mean trajectory, $\mathbf{q}(t$), we then have $(d/dt)ln[\rho(\mathbf{q}(t),t)]=-\nabla\cdot\mathbf{v}|_{\mathbf{q}=\mathbf{q}(t)}$.
Solving this gives $\rho(\mathbf{q}(t),t)=\rho_{0}(\mathbf{q}_{0})exp[-\int_{0}^{t}\left(\nabla\cdot\mathbf{v}\right)|_{\mathbf{q}=\mathbf{q}(t')}dt']$,
which implies that if $\rho_{0}(\mathbf{q}_{0})>0$, then $\rho(\mathbf{q}(t),t)>0$
for all times. Correspondingly, from $\rho(\mathbf{q}(t),t)$ we obtain
$R(\mathbf{q}(t),t)=R_{0}(\mathbf{q}_{0})-(\hbar/2)\int_{0}^{t}\left(\nabla\cdot\mathbf{v}\right)|_{\mathbf{q}=\mathbf{q}(t')}dt'$,
which never becomes undefined if $R_{0}(\mathbf{q}_{0})$ is not undefined.} Indeed, if the phase field would not undergo the discontinuous jump
at a nodal point, then this would imply that there are mean trajectories
near nodes for which the actual particle does not have a well-defined
mean phase, thereby contradicting our hypothesis. By contrast, for
the classical \emph{zbw} phase field, there is no reason for it to
be undefined at nodal regions since there is no nonlinear coupling
to the (inverse of the) probability density. Rather, the fact that
the classical phase field also satisfies a condition of the form (141)
implies that it changes discontinuously across a discontinuity in
the external potential, \emph{V}, and takes discrete values for changes
along a closed loop \emph{L} encircling the discontinuity in \emph{V}
(as demonstrated for the hydrogen-like atom in Appendix B ).

We thus have a formulation of ZSM in external fields that avoids the
Wallstrom criticism and is ready to be applied to the central potential
example considered in section 3.

\subsection{The central potential revisited}

With ZSM in hand, we can now return to the central potential example
considered by Wallstrom, and show how ZSM gives the same result as
quantum mechanics.

For the effective central potential, $V_{a}(\boldsymbol{\mathrm{r}})=V(\boldsymbol{\mathrm{r}})+a/r^{2}$,
we found that the HJM equations implied $\mathbf{v}'_{a}=\mathbf{v}{}_{a}\sqrt{2ma/\hbar^{2}+1}$
and $\mathbf{u}'_{a}=\mathbf{u}{}_{a}$, where $\mathbf{v}{}_{a}=\left(\hbar/mr\right)\hat{\varphi}$.
The problem in standard NYSM was that the constant $a$ could take
any positive real value, making $\mathbf{v}'_{a}$ not quantized.
By contrast, in the quantum mechanical version, $\mathrm{m}=\sqrt{2ma/\hbar^{2}+1}$
would be integral due to the single-valuedness condition on $\psi_{\mathrm{m}}$.

In the ZSM version of this problem, the \emph{zbw} phase field in
the lab frame, $S_{a}=\hbar\varphi$, satisfies 
\begin{equation}
\oint\frac{dS_{a}}{d\varphi}d\varphi=\oint\hbar d\varphi=\mathrm{m}h,
\end{equation}
as a consequence of the reasoning used in section 5.2. Accordingly,
for the effective \emph{zbw} phase field, $S'_{a}=\hbar\sqrt{2ma/\hbar^{2}+1}\varphi=\hbar\varphi'$,
we will also have 
\begin{equation}
\oint\hbar\sqrt{2ma+1}d\varphi=\oint\hbar d\varphi'=\mathrm{m}h,
\end{equation}
where $\mathrm{m}=\sqrt{2ma/\hbar^{2}+1}$ is integral. So ZSM predicts
quantized energy-momentum in the central potential case, in accordance
with quantum mechanics.

\section{Conclusion}

To answer Wallstrom's criticism, we first developed a classical \emph{zbw}
model (based on the earlier models of de Broglie and Bohm) which implies
a quantization condition reminiscent of the Bohr-Sommerfeld-Wilson
condition. We did this excluding and including interactions with external
fields, and formulated the classical Hamilton-Jacobi statistical mechanics
of each case. We then extended this model to Nelson-Yasue stochastic
mechanics - which we termed zitterbewegung stochastic mechanics (ZSM)
- and showed, using the same two cases, that it allows us to recover
the Schrödinger equation for single-valued wave functions with (in
general) multi-valued phases. Finally, we showed that ZSM works for
the concrete case of a two-dimensional central potential.

In Part II, our approach will be generalized to the case of many \emph{zbw}
particles, excluding and including (external and inter-particle) field
interactions, the latter of which turns out to be a non-trivial task.
We will also: (i) elaborate on the beables of ZSM, (ii) assess the
plausibility and generalizability of the \emph{zbw} hypothesis, and
(iii) compare ZSM to other (previously) proposed answers to Wallstrom's
criticism.

\section{Acknowledgments}

It is a pleasure to acknowledge helpful discussions with Guido Bacciagaluppi,
Dieter Hartmann, and Herman Batelaan. I am also gateful to Guido for
carefully reading an earlier draft of this paper and making useful
suggestions for improvements. Lastly, I thank Mike Towler for inviting
me to talk on an earlier incarnation of this work at his 2010 de Broglie-Bohm
research conference in Vallico Sotto, Tuscany, Italy.

\appendix
%dummy comment inserted by tex2lyx to ensure that this paragraph is not empty

\section{Proof of the Stochastic Variational Principle }

Following Yasue's presentation \cite{Yasue1981a}, let $\mathbf{q}'(t)=\mathbf{q}(t)+\delta\mathbf{q}(t)$
be a variation of the sample path $\mathbf{q}(t)$, with end-point
constraints $\delta\mathbf{q}(t_{i})=\delta\mathbf{q}(t_{f})=0$.
Let us also assume, for the sake of generality, that the particle
has charge \emph{e} and couples to the external magnetic vector potential,
$\mathbf{A}_{ext}(\mathbf{q}(t),t)$, as well as the external electric
scalar potential, $\Phi_{e}(\mathbf{q}(t),t)$. Then the condition
\begin{equation}
\begin{aligned}J & =\mathrm{E}\left[\int_{t_{i}}^{t_{f}}\left\{ \frac{1}{2}\left[\frac{1}{2}m\mathbf{b}(\mathbf{q}(t),t)^{2}+\frac{1}{2}m\mathbf{b}_{*}(\mathbf{q}(t),t)^{2}\right]+\frac{e}{c}\mathbf{A}_{ext}(\mathbf{q}(t),t)\cdot\mathbf{v}(\mathbf{q}(t),t)-e\Phi_{e}(\mathbf{q}(t),t)\right\} dt\right]\\
 & =\mathrm{E}\left[\int_{t_{i}}^{t_{f}}\left\{ \frac{1}{2}\left[\frac{1}{2}m\left(D\mathbf{q}(t)\right)^{2}+\frac{1}{2}m\left(D_{*}\mathbf{q}(t)\right)^{2}\right]+\frac{e}{c}\mathbf{A}_{ext}\cdot\mathbf{v}-e\Phi_{e}\right\} dt\right]=extremal,
\end{aligned}
\end{equation}
where $\mathrm{E}\left[...\right]$ is the absolute expectation, is
equivalent to the variation, 
\begin{equation}
\delta J(\mathbf{q})=J(\mathbf{q}')-J(\mathbf{q}),
\end{equation}
up to first order in $||\delta\mathbf{q}(t)||$. So (146) gives 
\begin{equation}
\begin{aligned}\delta J & =\mathrm{E}\left[\int_{t_{i}}^{t_{f}}\left\{ \left[\frac{1}{2}m\left(D\mathbf{q}(t)\cdot D\delta\mathbf{q}(t)+D_{*}\mathbf{q}(t)\cdot D_{*}\delta\mathbf{q}(t)\right)\right]\right.\right.\\
 & \left.\left.+\frac{e}{c}\mathbf{A}_{ext}\cdot\frac{1}{2}\left(D\delta\mathbf{q}(t)+D_{*}\delta\mathbf{q}(t)\right)+\frac{e}{c}\left(\delta\mathbf{q}(t)\cdot\nabla\mathbf{A}_{ext}\right)\frac{1}{2}\left(D\mathbf{q}(t)+D_{*}\mathbf{q}(t)\right)-e\nabla\Phi_{e}\cdot\delta\mathbf{q}(t)\right\} |_{\mathbf{q}=\mathbf{q}(t)}dt\right],
\end{aligned}
\end{equation}
where we note that $\mathbf{v}=\frac{1}{2}\left(D+D_{*}\right)\mathbf{q}(t)$
and is constrained by Eq. (10). Now, observing that for an arbitrary
function, $f(\mathbf{q}(t),t)$, we have 
\begin{equation}
\mathrm{E}\left[\int_{t_{i}}^{t_{f}}\left[f(\mathbf{q}(t),t)D\delta\mathbf{q}(t)\right]dt\right]=-\mathrm{E}\left[\int_{t_{i}}^{t_{f}}\left[\delta\mathbf{q}(t)D_{*}f(\mathbf{q}(t),t)\right]dt\right],
\end{equation}
and 
\begin{equation}
\mathrm{E}\left[\int_{t_{i}}^{t_{f}}\left[f(\mathbf{q}(t),t)D_{*}\delta\mathbf{q}(t)\right]dt\right]=-\mathrm{E}\left[\int d^{3}\mathbf{q}\rho\int_{t_{i}}^{t_{f}}\left[\delta\mathbf{q}(t)Df(\mathbf{q}(t),t)\right]dt\right],
\end{equation}
and 
\begin{equation}
\frac{1}{2}\left(D+D_{*}\right)f(\mathbf{q}(t),t)=\left\{ \frac{\partial}{\partial t}+\frac{1}{2}\left[D\mathbf{q}(t)+D_{*}\mathbf{q}(t)\right]\cdot\nabla\right\} f(\mathbf{q},t)|_{\mathbf{q}=\mathbf{q}(t)},
\end{equation}
we then obtain 
\begin{equation}
\begin{aligned}\delta J & =\mathrm{E}\left[\int_{t_{i}}^{t_{f}}\left\{ \frac{m}{2}\left[D_{*}D+DD_{*}\right]\mathbf{q}(t)\right.\right.\\
 & \left.\left.-\frac{e}{c}\mathbf{v}\times\left(\nabla\times\mathbf{A}_{ext}\right)+\frac{e}{c}\frac{\partial\mathbf{A}_{ext}}{\partial t}+e\nabla\Phi_{e}\right\} |_{\mathbf{q}=\mathbf{q}(t)}\delta\mathbf{q}(t)dt\right]+\vartheta(||\delta\mathbf{q}||).
\end{aligned}
\end{equation}
From the variational constraint (145-46), it follows that the first-order
variation of $J$ must be zero for arbitrary sample-wise variation
$\delta\mathbf{q}(t)$. Moreover, since the expectation is a positive
linear functional, we will have the equation of motion 
\begin{equation}
\frac{m}{2}\left[D_{*}D+DD_{*}\right]\mathbf{q}(t)=-e\left[\nabla\Phi_{e}+\frac{1}{c}\frac{\partial\mathbf{A}_{ext}}{\partial t}\right]|_{\mathbf{q}=\mathbf{q}(t)}+\frac{e}{c}\mathbf{v}\times\left(\nabla\times\mathbf{A}_{ext}\right)|_{\mathbf{q}=\mathbf{q}(t)}
\end{equation}
for each time $t$ $\in$ $\left[t_{i},t_{f}\right]$ with probability
one.

\section{\textcolor{black}{Classical Zitterbewegung in the Central Potential}}

Suppose that the non-relativistic \emph{zbw} particle in the lab frame
is moving in a circular orbit about some central potential, $V(\mathbf{r}),$
where $\mathbf{r}$ is the radius of the orbit. In this case, for
the spherical coordinates $(r,\alpha,\beta)$, $r$ is fixed, $\alpha$
is varies with time, and $\beta$ has the constant value $\pi/2$,
giving translational velocities $\mathbf{v}_{r}=\dot{r}=0$ (and we
require $\ddot{r}=0$), $\mathbf{v}_{\alpha}=r\dot{\alpha}$, and
$\mathbf{v}_{\beta}=r\dot{\beta}sin\alpha=0$. The $v\ll c$ approximated
\emph{zbw} phase change in the lab frame is then 
\begin{equation}
\begin{aligned}\delta\theta(\alpha(t),t) & =\left(\omega_{c}+\omega_{\alpha}+\kappa(\mathbf{r})\right)\delta t-\frac{v_{\alpha}r\delta\alpha(t)}{c^{2}}\\
 & =\frac{\omega_{c}}{mc^{2}}\left[\left(mc^{2}+\frac{p_{\alpha}^{2}}{2mr^{2}}+V(\mathbf{r})\right)\delta t-v_{\alpha}r\delta\alpha(t)\right],
\end{aligned}
\end{equation}
where $p_{\alpha}=mr^{2}\dot{\alpha}$. Because the total energy of
the system is constant, integrating this gives 
\begin{equation}
\theta=\frac{\omega_{c}}{mc^{2}}\left[\left(mc^{2}+\frac{p_{\alpha}^{2}}{2mr^{2}}+V(\mathbf{r})\right)t-p_{\alpha}\alpha(t)\right]+C,
\end{equation}
or 
\begin{equation}
\begin{aligned}S & =p_{\alpha}\alpha(t)-\left(mc^{2}+\frac{p_{\alpha}^{2}}{2mr^{2}}+V(\mathbf{r})\right)t+C\\
 & =p_{\alpha}\alpha(t)-Et+C.
\end{aligned}
\end{equation}
Incidentally, we could have also obtained (155) by starting with the
non-relativistic Lagrangian 
\begin{equation}
L(\mathbf{\alpha}(t),t)=\frac{1}{2}mr{}^{2}\dot{\alpha}(t)^{2}-V(\mathbf{r})-mc^{2},
\end{equation}
and using the Legendre transformation, 
\begin{equation}
E=p_{\alpha}\dot{\mathbf{\alpha}}-L=\frac{p_{\alpha}^{2}}{2mr^{2}}+V(\mathbf{r})+mc^{2},
\end{equation}
to get 
\begin{equation}
S=\int Ldt+C=\int\left(p_{\alpha}\dot{\mathbf{\alpha}}-E\right)dt+C=p_{\alpha}\alpha-Et+C.
\end{equation}
Clearly (155) satisfies the classical Hamilton-Jacobi equation 
\begin{equation}
-\frac{\partial S}{\partial t}=\frac{1}{2mr^{2}}\left(\frac{\partial S}{\partial\alpha}\right)^{2}+V(\mathbf{r}),
\end{equation}
where $-\partial S/\partial t=E$ and $\partial S/\partial\alpha=\mathbf{p}_{\alpha}=\mathbf{L}_{\alpha}$,
the latter being the constant angular momentum of the particle in
the $\hat{z}$-direction.

Because the \emph{zbw} oscillation is simply harmonic and the phase
is a well-defined function of the particle position, the change in
$S$ will now be quantized upon fixed time integration around a closed
(circular) orbit \emph{L}. In other words, we will have 
\begin{equation}
\oint_{L}p_{\alpha}\delta\alpha=2\pi mv_{\alpha}r\mathrm{=}nh,
\end{equation}
or 
\begin{equation}
L_{\alpha}=mv_{\alpha}r=n\hbar,
\end{equation}
where $n$ is an integer. From (161) and the force balance equation
(assuming a Coulomb force), $mv_{\alpha}^{2}/r=(1/4\pi\epsilon_{0})e^{2}/r^{2}$,
it follows that the radius is quantized as 
\begin{equation}
r_{n}=\frac{4\pi\epsilon_{0}\hbar^{2}}{m_{e}e^{2}}n^{2},
\end{equation}
where for $n=1$, (162) gives the Bohr radius. Inserting (162) into
the force balance equation and recognizing that $E=V/2$, we then
obtain the quantized energy states 
\begin{equation}
E_{n}=\frac{E_{1}}{n^{2}},
\end{equation}
where $E_{1}=-e^{2}/8\pi\epsilon_{0}r_{1}=-13.6eV$ is precisely the
magnitude of the ground state energy of the Bohr hydrogen atom.

We wish to emphasize that, whereas Bohr simply assumed a condition
equivalent to (160) in order to stabilize the electron's circular
orbit in the classical hydrogen atom, we \textit{\textcolor{black}{obtained}}
(160) just from the zitterbewegung hypothesis in the particle's instantaneous
translational rest frame combined with the usual Lorentz transformation.
In other words, in Bohr's model, (160) is imposed ad hoc while in
our model it arises as a direct consequence of a relativistic (\emph{zbw})
constraint on the particle's motion.

\bibliographystyle{unsrt}
\bibliography{PhDthesisRefscopy}

\end{document}